\documentclass[aps,preprint,floats,nofootinbib,superscriptaddress]{revtex4}

\usepackage{graphicx}

\begin{document}

\preprint{
\hfill
\begin{minipage}[t]{3in}
\begin{flushright}
\vspace*{-0.3in}
NUB--3257--Th--05\\
MADPH--05--1435\\
FERMILAB--PUB--05--373--A\\
hep-ph/0508312
\end{flushright}
\end{minipage}
}

\hfill$\vcenter{\hbox{}}$

\thispagestyle{empty}

\title{Exotic Neutrino Interactions at the Pierre Auger Observatory}

\author{Luis Anchordoqui}
\affiliation{Department of Physics, Northeastern University,\\ 
Boston, MA 02115, USA\\}
 
\author{Tao Han}
\affiliation{Department of Physics,
University of Wisconsin,\\ Madison, WI 53706, USA\\}

\author{Dan Hooper}
\affiliation{Particle Astrophysics Center, Fermilab,\\ 
P.O. Box 500, Batavia, IL 60510, USA}
\affiliation{Astrophysics, University of Oxford,\\ Oxford OX1 3RH, UK\\}

\author{Subir Sarkar}
\affiliation{Rudolf Peierls Centre for Theoretical Physics, University
of Oxford,\\ 
Oxford OX1 3NP, UK}

\date{\today}

\begin{abstract}
\bigskip\noindent  
The Pierre Auger Observatory for cosmic rays provides a laboratory for
studying fundamental interactions at energies well beyond those
available at colliders. In addition to hadrons or photons, Auger is
sensitive to ultra-high energy neutrinos in the cosmic radiation and
models for new physics can be explored by observing neutrino interactions
at center-of-mass energies beyond the TeV scale. By comparing the rate
for quasi-horizontal, deeply penetrating air showers triggered by all
types of neutrinos with the rate for slightly upgoing showers
generated by Earth-skimming tau neutrinos, any deviation of the
neutrino-nucleon cross-section from the Standard Model expectation can
be constrained. We show that this can test models of low-scale quantum
gravity (including processes such as Kaluza-Klein graviton exchange,
microscopic black hole production and string resonances), as well as
non-perturbative electroweak instanton mediated processes. Moreover,
the observed ratios of neutrino flavors would severely constrain the
possibility of neutrino decay.
\end{abstract}

\maketitle

\section{Introduction}

The Pierre Auger Observatory is the largest cosmic ray detector in the
world ~\cite{Abraham:2004dt}. Currently under construction at
Malarg\"ue (Argentina), it has begun taking data and already
accumulated an exposure comparable to previous experiments such as
AGASA and HiRes~\cite{auger}.

In addition to studying the highest energy cosmic rays, Auger is also
capable of observing ultra-high energy cosmic
neutrinos~\cite{augersim}. At present, the AMANDA telescope at the
South Pole holds the record for the most energetic neutrino
interactions observed~\cite{amanda}; these events have energies up to
$\sim 10^{5}$~GeV and are consistent with the predicted spectrum of
atmospheric neutrinos. Auger, by contrast, is expected to detect
neutrinos with energies above $\sim 10^{8}$~GeV. The ability to study
neutrino interactions at such high energies will open a unique window
on possible physics beyond the Standard Model (SM) of strong and
electroweak interactions.

A variety of models have been proposed in which neutrino interactions
become substantially modified at very high energies, the most
interesting being models of low scale quantum gravity (involving the
exchange of Kaluza-Klein (KK) gravitons, production of microscopic
black holes, and excitation of TeV-scale string resonances), and
models featuring non-perturbative electro-weak instanton induced
interactions.

Moreover the neutrino flavor ratios predicted by the standard
oscillation phenomenology can be modified in propagation over
cosmological distances if processes such as neutrino decay occur.
Auger is expected to detect the `cosmogenic' neutrino flux from
interactions of extragalactic ultra-high energy cosmic rays with the
cosmic microwave background. Thus, it will be sensitive to such
effects, being capable of measuring the flux of ultra-high energy tau
neutrinos in addition to the overall neutrino flux.

In this article, we explore the relevant phenomenology and quantify
the sensitivity of Auger to such new physics.\footnote{For a review of
exotic neutrino interactions and their signatures in high-energy
cosmic neutrino telescopes such as IceCube, see
Ref.~\cite{Han:2004kq}.} In Sec.~\ref{detector}, we describe the Auger
experiment and its ability to detect quasi-horizontal neutrino-induced
air showers, as well as up-going showers induced by Earth-skimming tau
neutrinos. In Sec.~\ref{fluxes} we discuss possible sources of cosmic
ultra-high energy neutrinos. In section~\ref{general}, we infer the
sensitivity of Auger to the neutrino-nucleon interaction cross-section
and to the flavour content of the ultra-high energy cosmic neutrino
flux. In Sec.~\ref{exotic} we consider specific models of physics
beyond the SM and their signatures in Auger. We present our
conclusions in Sec.~\ref{conclusions}.

\section{Ultra-High Energy Neutrinos at Auger}
\label{detector}

\subsection{The Pierre Auger Observatory}

Auger is a hybrid ultra-high energy cosmic ray detector, with a ground
array of water Cerenkov detectors sampling air shower particles,
overlooked by air fluorescence detector telescopes which observe the
longitudinal development of the showers~\cite{Abraham:2004dt}. When
completed in 2005--06, the Southern hemisphere site will have 1600
detectors on the ground covering 3000 km$^2$, and 4 fluorescence
telescopes having 6 detectors each. A similar facility has been
proposed for a Northern hemisphere site in Colorado (USA). In our
calculations we consider only a single site and focus on the ground
array.

\subsection{Quasi-Horizontal, Deeply Penetrating Showers}
\label{QH}

At sufficiently high energies cosmic neutrinos can trigger atmospheric
air showers similar to those due to high energy cosmic rays (hadrons
or photons). However, unlike ordinary cosmic ray showers which are
initiated near the top of the atmosphere, those generated by neutrinos
can be initiated at any depth since the interaction cross-section is
much smaller, hence the probability of interaction per unit length is
approximately constant. Neutrino induced showers can thus be
distinguished from cosmic ray showers by requiring that they be {\em
deeply penetrating}. This is most useful for distinguishing the two
kinds of showers, because within $\sim20^{\circ}$ of the horizon the
electromagnetic component of hadron-induced showers is completely
absorbed before reaching the detector.

\begin{figure}[tbh]
\centering\leavevmode
\mbox{
\includegraphics[width=3.5in,angle=90]{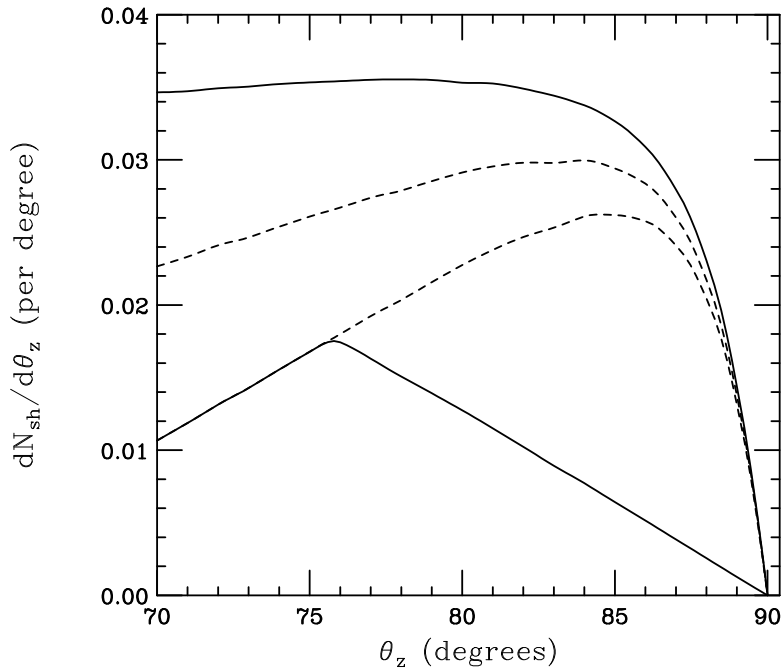}
}
\caption{Angular distribution of neutrino-induced showers expected to
be observed by the Auger ground array with different selection
criteria. The upper solid line indicates all showers generated by
cosmic neutrinos over the range of zenith angles shown, while the
upper and lower dashed lines correspond to the cases when the shower
is initiated at a depth exceeding 1000 and 2000 g/cm$^2$,
respectively. The lower solid line corresponds to the case when the
shower is initiated at a depth exceeding 2000 g/cm$^2$ {\it and}
within 2000 g/cm$^2$ from the detector --- this last set of cuts is
adopted in our calculations. For the four cases shown, the fraction of
events which survive the various cuts are 1.0, 0.80, 0.60 and 0.33,
respectively. We have adopted a neutrino spectrum $\propto E^{-2},$
saturating the Waxman-Bahcall flux bound, see Eq.~(\ref{WB}).}
\label{zenith}
\end{figure}

The rate of neutrino-induced showers expected to be observed in an
experiment such as Auger can be written as
\begin{eqnarray}
\frac{N_{\rm events}}{\Delta T_{\rm obs}} & = &  
2 \pi N_{\rm A}\,\int{\rm d}E_\nu\,\int^1_0 {\rm d}y 
\frac{{\rm d}\sigma_{\nu N}}{{\rm d}y}(E_\nu) 
\,\int{\rm d}\cos\theta_{\rm z}\,A_{\perp}(\cos\theta_{\rm z}) \nonumber \\
 & \times & \int^{X_{\rm ground}}_{X_{\rm min}} \, 
{\rm d}X \, P[E_{\rm sh},\cos\theta_{\rm z}, X] \, 
\frac{{\rm d}N_\nu}{{\rm d}E_\nu}(E_\nu) \,\,,
\label{qh}
\end{eqnarray}
where $\Delta T_{\rm{obs}}$ is the observation time, $N_{\rm A}$ is
Avogadro's number, ${\rm d}\sigma_{\nu N}/{\rm d}y$ is the
differential neutrino-nucleon cross-section, $y$ is the inelasticity,
$\theta_{\rm z}$ is the zenith angle, $A_{\perp}$ is the cross
sectional area of the experiment as seen from a given zenith angle,
$X$ is the atmospheric depth of the interaction (the atmospheric mass
per unit area), $E_{\rm sh}$ is the total energy dissipated in the
shower, and ${\rm d}N_{\nu}/{\rm d}E_{\nu}$ is the incoming neutrino
flux.  The function $P[E_{\rm{sh}},\theta_{\rm z}, X]$ is the
probability of the experiment detecting a shower created at an
atmospheric depth $X$ of energy $E_{\rm{sh}}$, at a zenith angle
$\theta_{\rm z}$. In order to ensure that such a shower can be
distinguished from one initiated by a hadron or photon primary, we
require that for the Auger ground array:
\begin{itemize}

\item The zenith angle, $\theta_{\rm z} > 70^{\circ}$, 

\item The neutrino penetrates at least 2000 g/cm$^2$ into the
atmosphere before interacting,

\item The interaction takes place within 2000 g/cm$^2$ of the detector.

\end{itemize}
This third requirement is included due to the difficulty in
reconstructing events beyond this range at Auger. At $\theta_{\rm
z}=70^{\circ}$, the total path length traversed before the shower hits
the Earth's surface is $X_{\rm{ground}} \approx 3000$ g/cm$^2$, hence
relatively little atmosphere is present in which a neutrino primary
can interact and be distinguished from an ordinary cosmic ray
shower. At zenith angles larger than 85$^{\circ}$ however, the slant
depth exceeds 10000 g/cm$^2$. In Fig.~\ref{zenith}, we show how these
selection criteria will affect the observed angular distribution of
quasi-horizontal showers in Auger.

\begin{figure}[tbh]
\centering\leavevmode
\mbox{
\includegraphics[width=3.5in,angle=90]{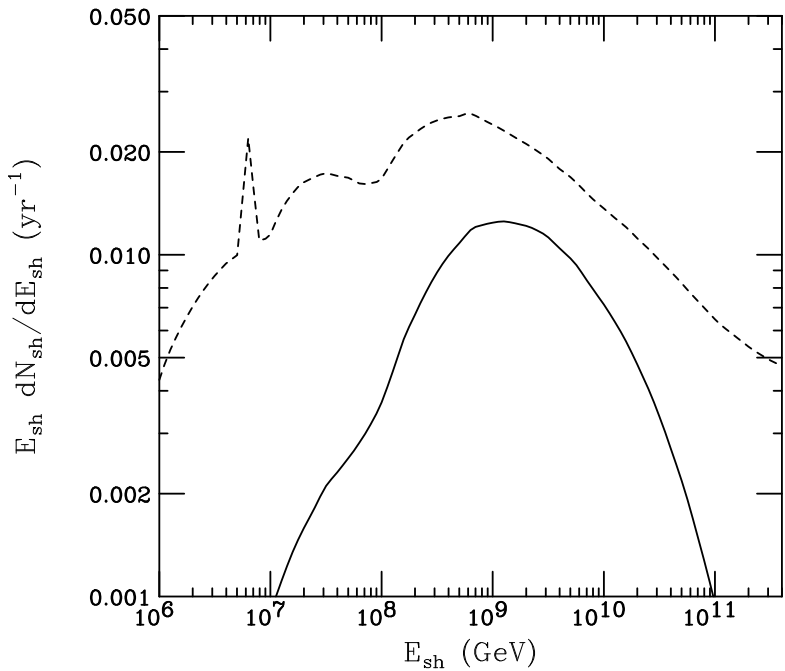}
}
\caption{Spectra of quasi-horizontal, deeply penetrating, neutrino
induced showers as would be seen by Auger with the cuts shown in
Fig.\ref{zenith}. The solid and dashed lines correspond to the
cosmogenic flux and the Waxman-Bahcall flux (see section~\ref{fluxes}
for details) which yield, respectively, 0.07 and and 0.22 events per
year. For the latter case, the `bumps' at $\sim10^{7.5}$ GeV and
$\sim10^{8.5}$ GeV correspond to NC and CC interactions respectively,
while the $W^-$ resonance can be seen at $6.3 \times 10^{6}$~GeV.}
\label{cossmqh}
\end{figure}

If the neutrino-nucleon cross-section is enhanced well above the SM
prediction at ultra-high energies, it is possible that this depth of
atmosphere will significantly attenuate the cosmic neutrino flux; to
account for this, an additional factor of ${\rm e}^{-N_{\rm A} \,
(X_{\rm{ground}}-2000\, \rm{g/cm}^2)\, \sigma_{\nu N}}$ should be
included~\cite{atten} in Eq.~(\ref{qh}). For SM interactions, this
factor is very nearly unity and can safely be neglected but it will be
important for some of the exotic models we will consider.

The neutrino-nucleon cross-section in Eq.~(\ref{qh}) describes both
charged current (CC) neutrino-quark scattering and neutral current
(NC) neutrino-quark scattering for which we adopt the cross-sections
given in Ref.~\cite{cross}. The energy of the shower produced depends
on the neutrino flavor and the type of
interaction~\cite{Anchordoqui:2002vb}.  Electron neutrinos undergoing
CC interactions produce a shower with both an electromagnetic and
hadronic component: $E_{\rm{sh, em}} = (1-y) E_{\nu}$, $E_{\rm{sh,
had}} = y E_{\nu}$. Muon neutrinos undergoing CC interactions, as well
as all neutrino flavors undergoing NC interactions, produce a hadronic
shower with an energy, $E_{\rm{sh, had}} = y E_{\nu}$.

Charged current interactions of tau neutrinos are somewhat more
complicated but more interesting. The tau lepton produced in the
initial CC neutrino interaction has a decay length of $L_{\tau}
\approx$ 50~m $\times \, (E_{\tau}/10^6~{\rm GeV})$. Thus, at
sufficiently high energies, the second hadronic shower from the tau
decay will be spatially separated and be identifiable as a ``double
bang'' event~\cite{doublebang}. Lacking a full-blown simulation, we
estimate that a separation of the two bangs by 10~km would be adequate
for definitive identification by the Auger ground array --- this
requires that the primary neutrino energy exceed $\sim 3\times
10^{9}$~GeV and that the first interaction occurs 50 km or more above
the ground (which is easily satisfied for neutrino induced showers
inclined over $80^\circ$). However, if the energy exceeds
$\sim10^{10}$~GeV, the tau lepton will hit the ground before
decaying. A more careful analysis of shower profiles as seen by the
Auger fluorescence detectors may allow significant acceptance for such
events over a somewhat broader energy range.

The scattering of electron flavor anti-neutrinos with electrons can
occur efficiently via the resonant exchange of a $W^-$
boson~\cite{Glashow:W} at a neutrino energy of $6.3 \times
10^{6}$~GeV. Although we include this process in our calculations, the
detector acceptance for showers at this energy is expected to be
rather low, thus this process is of only marginal importance.

To accurately determine the probability $P[E_{\rm{sh}},\cos
\theta_{\rm z}, X]$ of the Auger ground array observing a shower with
a given energy, zenith angle and initiated at a given depth, a
detailed detector simulation is required, which is beyond the scope of
this study. To make a reasonable estimate we have modelled the energy
dependence such that we reproduce the acceptances found through the
simulations performed in Ref.~\cite{augersim}. The probability
function we arrive at is of order unity for shower energies of ${\cal
O}(10^{12})$ GeV, decreases slowly down to energies of ${\cal
O}(10^9)$~GeV, and then falls rapidly at lower energies. We treat
hadronic and electromagnetic showers separately as in
Ref.~\cite{augersim}; in the case of a mixed electromagnetic-hadronic
shower, we treat it as two separate showers for the purpose of
estimating the probability of detection. In Fig.~\ref{cossmqh} we plot
the spectrum of quasi-horizontal, deeply penetrating, neutrino induced
showers as would be seen by the Auger ground array, for two choices
of the ultra-high energy cosmic neutrino spectrum.

\subsection{Earth Skimming Tau Neutrinos}

A second class of neutrino events potentially observable at Auger is
generated by tau neutrinos which interact while skimming the Earth's
surface \cite{tauevents}. Such interactions can generate tau leptons
which escape the Earth and produce a slightly upgoing hadronic shower
when they decay in the atmosphere.\footnote{Even when the tau lepton
decays in the Earth, regeneration effects~\cite{regeneration} can
extend the effective range, such that another tau lepton emerges and
decay in the atmosphere.} This does not happen for electron neutrinos
since the electrons produced in CC interactions are invariably
absorbed in the Earth. For muon neutrinos, the produced muon can
escape the Earth, but will not decay in the atmosphere since the decay
length is $\agt 10^{8}$ times longer than for a tau.

\begin{figure}[tbh]
\centering\leavevmode
\mbox{
\includegraphics[width=3.5in,angle=90]{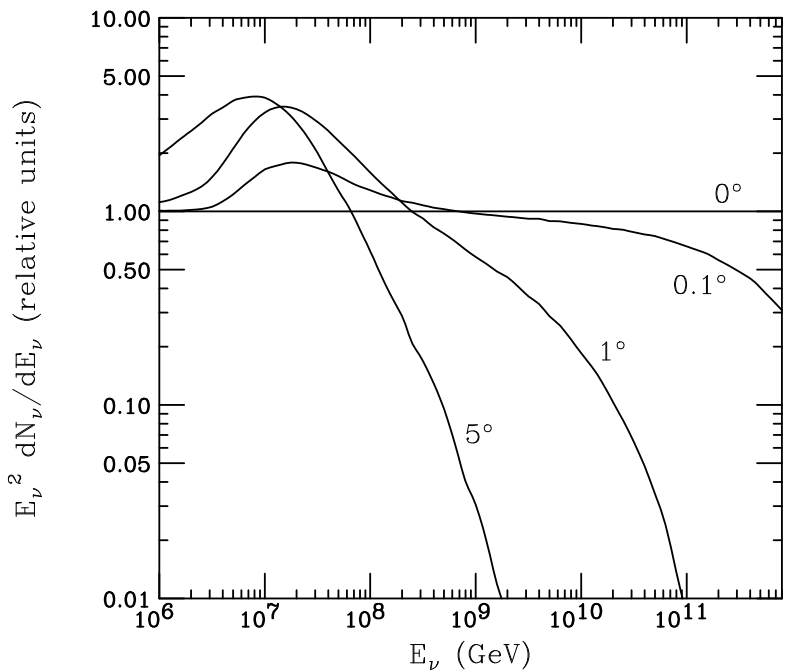}
}
\caption{Effect of interactions in the Earth on cosmic tau neutrinos
with a spectrum $\propto E^{-2}_{\nu}$ extending up to
$10^{12}$~GeV. The horizontal line is the unmodified spectrum and the
other lines are for incoming angles 0.1$^{\circ}$, 1$^{\circ}$ and
5$^{\circ}$ degrees below the horizon.}
\label{tauprop}
\end{figure}

When a tau lepton is generated in the Earth, it loses energy via
electromagnetic processes at a rate per unit length of~\cite{tauevents}
\begin{equation}
\frac{dE_{\tau}}{dx} \approx -\alpha -\beta \, E_{\tau},
\label{tauloss}
\end{equation}
where $\alpha$= 0.002 GeV cm$^2$/g and $\beta=6 \times 10^{-7}$
cm$^2$/g. At very high energies, this will often dramatically reduce
the energy before the tau is able to decay. At moderate energies, this
has little effect over a single decay length. In Fig.~\ref{tauprop} we
show the effect of interactions for tau neutrino `Earth skimmers'. For
incoming zenith angles only slightly below the horizon, the spectrum
is not suppressed until above $\sim 10^9$~GeV, while there is a
noticeable pile-up near $10^{7}$~GeV.

\begin{figure}[tbh]
\centering\leavevmode
\mbox{
\includegraphics[width=3.5in,angle=90]{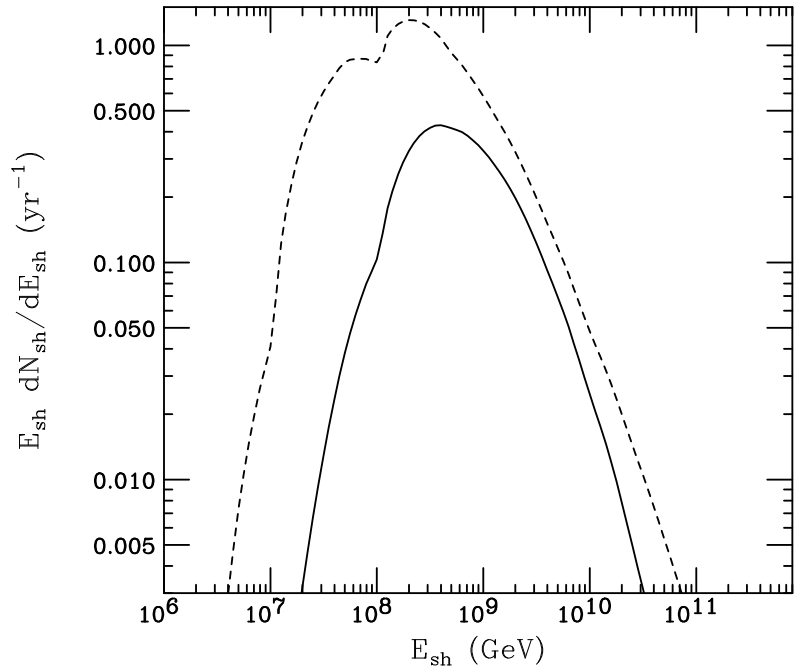}
}
\caption{The spectrum of Earth skimming, tau neutrino induced showers
as would be seen by Auger. The solid and dashed lines are,
respectively, for the cosmogenic neutrino flux and the Waxman-Bahcall
flux, which would yield 1.3 and 4.8 events per year in Auger.}
\label{cossmtau}
\end{figure}

Tau leptons produced in CC interactions near the Earth's surface can
occasionally escape the Earth before decaying, and thus produce a
hadronic shower which is potentially observable by Auger. We have
calculated the spectrum of tau leptons escaping the Earth's surface by
Monte Carlo using the energy loss rate of Eq.~(\ref{tauloss}). To
calculate the probability of given tau neutrino induced shower being
detected by Auger, we have used the same probabilities as employed in
the case of quasi-horizontal showers. In addition to this function,
however, we require that the shower be initiated at a height such that
the shower is still able to be detected (for the details of this
aspect of the calculation, see Ref.~\cite{zas}). This is particularly
important for very high energy tau leptons which can escape the
Earth's atmosphere before decaying~\cite{tauevents}. The Andes
mountains near Auger's southern site are also a possible target for
tau neutrinos; however, we have not included this effect in our
calculations as the overall correction is less than
10\%~\cite{andes}. In Fig.~\ref{cossmtau} we show the spectrum of
Earth skimmers as would be seen by Auger.

\section{Ultra-High Energy Cosmic Neutrino Fluxes}
\label{fluxes}

Ultra-high energy neutrinos may be produced in a wide range of
astrophysical sources. In this section, we briefly discuss some of
these possibilities.

Interactions of ultra-high energy cosmic ray protons propagating over
cosmological distances with the cosmic microwave background generates
a cosmogenic flux of neutrinos~\cite{Berezinsky:1969} through the
decay of charged pions produced in $p \gamma$ interactions~\cite{gzk},
which should also result in a suppression of the cosmic ray spectrum
above the `GZK cutoff': $E_{\rm GZK} \sim 5 \times 10^{10}$ GeV. The
intermediate state of the reaction $p \gamma_{\rm CMB} \to N \pi$ is
dominated by the $\Delta^+$ resonance, because the $n$ decay length is
smaller than the nucleon mean free path on the relic photons.  Hence,
there is roughly an equal number of $\pi^+$ and $\pi^0$. Gamma rays,
produced via $\pi^0$ decay, subsequently cascade electromagnetically
on the cosmic radiation fields through $e^+ e^-$ pair production
followed by inverse Compton scattering.  The net result is a pile up
of $\gamma$ rays at GeV energies, just below the threshold for further
pair production.  On the other hand, each $\pi^+$ decays to 3
neutrinos and a positron.  The $e^+$ readily loses its energy through
synchrotron radiation in the cosmic magnetic fields.  The neutrinos
carry away about 3/4 of the $\pi^+$ energy, and therefore the energy
in cosmogenic neutrinos is about 3/4 of the one produced in
$\gamma$-rays.

The normalisation of the neutrino flux depends critically on the
cosmological evolution of the cosmic ray sources and on their proton
injection spectra~\cite{Yoshida:pt,engel}. It also depends on the
assumed spatial distribution of sources; for example, relatively local
objects, such as sources in the Virgo cluster~\cite{Hill:1985mk},
would dominate the high energy tail of the neutrino spectrum.  Another
source of uncertainty in the cosmogenic neutrino flux is the energy at
which there is a transition from galactic to extragalactic cosmic rays
as inferred from a change in the spectral slope. While Fly's Eye
data~\cite{Bird:1993yi} seem to favour a transition at $10^{10}$~GeV,
a recent analysis of the HiRes data~\cite{Bergman:2004bk} points to a
lower value of $\sim 10^{9}$~GeV. This translates into rather
different proton luminosities at the sources~\cite{lowcrossover} and
consequently different predictions for the expected flux of
neutrinos~\cite{Fodor:2003ph}.  A fourth source of uncertainty in the
cosmogenic flux is the chemical composition --- if ultra-high energy
cosmic rays are heavy nuclei rather than protons the corresponding
cosmogenic neutrino flux may be somewhat
reduced~\cite{heavycosmogenic}. Throughout this paper, we will adopt
the cosmogenic neutrino spectrum as calculated in Ref.~\cite{engel}.

\begin{figure}[tbh]
\centering\leavevmode
\mbox{
\includegraphics[width=3.5in,angle=90]{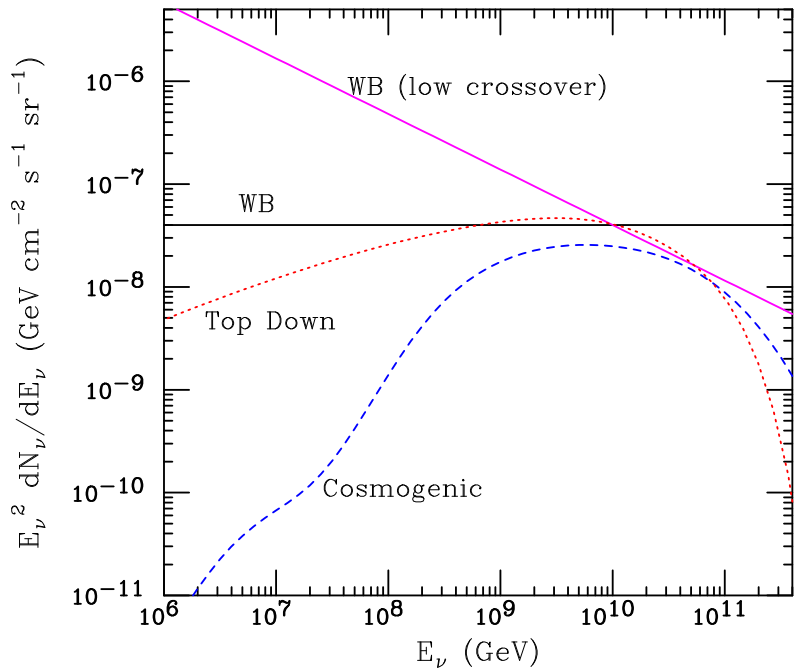}
}
\caption{Different possibilities for the ultra-high energy cosmic
neutrino spectrum. The solid horizontal line (WB) corresponds to the
Waxman and Bahcall bound \cite{wb} assuming proton-proton
interactions. The more rapidly falling solid line (WB (low crossover))
is obtained by the same argument but assuming a lower energy
transition between galactic and extragalactic cosmic rays
\cite{lowcrossover}. The dotted line (Top Down) is the predicted flux
in models where the highest energy cosmic rays arise from the decays
of superheavy dark matter particles to many body states
\cite{dreesneutrinos}. Finally, the dashed line (Cosmogenic) is the
spectrum of neutrinos produced in the intergalactic propagation of
ultra-high energy protons \cite{engel}. In all cases, the curves show
the sum of neutrinos and anti-neutrinos of all flavors.}
\label{fluxfig}
\end{figure}

In addition to being produced in the propagation of ultra-high energy
cosmic rays, neutrinos are also expected to be generated in their
sources, such as gamma-ray bursts or active galactic nuclei
\cite{review}. Although the details of the relationship between the
cosmic ray spectrum and the cosmic neutrino spectrum are model
dependent, some rather general arguments can be applied. In
particular, Waxman and Bahcall~\cite{wb} have shown that for compact,
cosmological sources, which are optically thin to proton-proton and
proton-photon interactions, an upper limit can be placed on the flux
of neutrinos. We will follow them in adopting a neutrino spectrum
arising from proton-proton collisions (with an inelasticity of 60\%):
\begin{equation}
E_\nu^2 {\rm d}N_\nu/{\rm d}E_\nu \lesssim 4 \times 10^{-8}~{\rm GeV} \, 
{\rm cm}^{-2} \, {\rm s}^{-1}\, {\rm sr}^{-1} \,,
\label{WB}
\end{equation}
summed over all flavors. After oscillations during propagation, one
finds at Earth a nearly identical flux of the three neutrino flavors
\cite{doublebang} with equal number of neutrinos and antineutrinos
\cite{equal}. If the shape of the neutrino spectrum is not an $E^{-2}$
power law, or if the other assumptions of the Waxman-Bahcall argument
are modified, this bound can be exceeded~\cite{mpr}. For example, if
their bound is evaluated under the assumption of a low galactic to
extragalactic crossover energy ($\sim 4 \times 10^8$ GeV rather than
the $\sim10^{10}$ GeV used by Waxman and Bahcall) a larger flux with a
steeper spectrum ($E^{-2.54}$) is obtained
\cite{lowcrossover}. Furthermore, sources which are optically thick
such that only neutrinos can escape (`hidden sources'), can easily
exceed this bound~\cite{hidden}).

Finally, if the highest energy cosmic rays are not accelerated in
distant astrophysical sources but are instead produced relatively
locally in the galactic halo in the decays of supermassive dark matter
particles, then significantly higher fluxes of ultra-high energy
photons and neutrinos will also be generated~\cite{topdown}. This
model was motivated by the AGASA observation that the cosmic ray
spectrum continues apparently without attenuation beyond $E_{\rm GZK}$
but at the same time the events are isotropically distributed on the
sky. Such events have not yet been seen by Auger, which has moreover
set a restrictive upper limit on the fraction of photons in the cosmic
ray flux \cite{augerphoton}. We have normalized the theoretical
expectations for the neutrino flux from QCD and electroweak
fragmentation in heavy particle decay as calculated in
Ref.~\cite{dreesneutrinos} by matching the flux of nucleons observed
by Auger, and checked that the photon limits are not violated.

In Fig.~\ref{fluxfig}, we plot the expected spectrum of ultra-high
energy cosmic neutrinos in the models discussed above and give the
corresponding event rates for Auger with standard QCD parton model
calculations in Table~\ref{fluxestable}. In this calculation we have
truncated the cosmic neutrino spectra above $10^{12}$ GeV --- this
choice has only a mild effect on the estimated rates.

\begin{table}[!ht]
\begin{tabular} {| c || c | c |} 
\hline
& Quasi-horizontal & Earth-skimming $\nu_{\tau}$ \\
\hline \hline
Cosmogenic & 0.067 & 1.3 \\
\hline
Waxman-Bahcall & 0.22 & 4.8 \\
\hline
Waxman-Bahcall (low crossover) & 2.1 & 35 \\
\hline
Top-Down & 0.16 & 4.1 \\
\hline
\end{tabular}
\caption{The number of neutrino induced events per year expected in
Auger for various choices of the ultra-high energy neutrino spectrum,
as shown in Figure~\ref{fluxfig}, calculated using the standard QCD
parton model cross-section.}
\label{fluxestable}
\end{table}

\section{Neutrino Physics with Auger}
\label{general}

\subsection{Prospects for Cross-Section Measurements}
\label{sigmameasure}

Deviations of the neutrino-nucleon cross-sections from the prediction
of the simple parton model~\cite{cross} can signal new physics beyond
the SM, but might alternatively be just due to saturation effects
which can substantially modify the parton density at small $x$
(i.e. small energy fractions)~\cite{Gribov:1984tu}.  These effects can
significantly reduce the total cross-section at high energies,
softening the power law behavior predicted by the simple `unscreened'
parton model toward compliance with the Froissart
bound~\cite{Kwiecinski:1990tb}.  By contrast, new physics such as
TeV-scale quantum gravity~\cite{Arkani-Hamed:1998rs,Randall:1999ee}
can enhance the neutrino interaction cross-sections.\footnote{It is
noteworthy that the neutrino-nucleon cross section can also be
enhanced in some supersymmetric models through direct channel
production of superpartner resonances~\cite{Carena:1998gd}.}  This has
been calculated in various different frameworks, e.g., arising from
exchange of Kaluza-Klein (KK)
gravitons~\cite{Nussinov:1998jt,Jain:2000pu}, black hole
production~\cite{Feng:2001ib}, and TeV-scale string
excitations~\cite{Domokos:1998ry}. In this section, we will discuss
the ability of Auger to measure deviations in the neutrino-nucleon
scattering cross-section from the SM prediction, without assuming any
particular interaction model.

The event rates for quasi-horizontal and Earth-skimming neutrinos have
different responses to the inelastic
cross-section~\cite{measureair}. The rate of quasi-horizontal showers
{\em rises} proportional to the cross-section (although if this
exceeds $\sim 10^{-28}$ cm$^2$, attenuation of the neutrino flux in
the upper atmosphere becomes significant~\cite{atten}). By contrast,
the rate of Earth skimming tau events is always {\em depleted} by an
enhanced neutrino-nucleon cross-section because of absorption in the
Earth.

\begin{figure}[tbh]
\centering\leavevmode
\mbox{
\includegraphics[width=1.7in,angle=90]{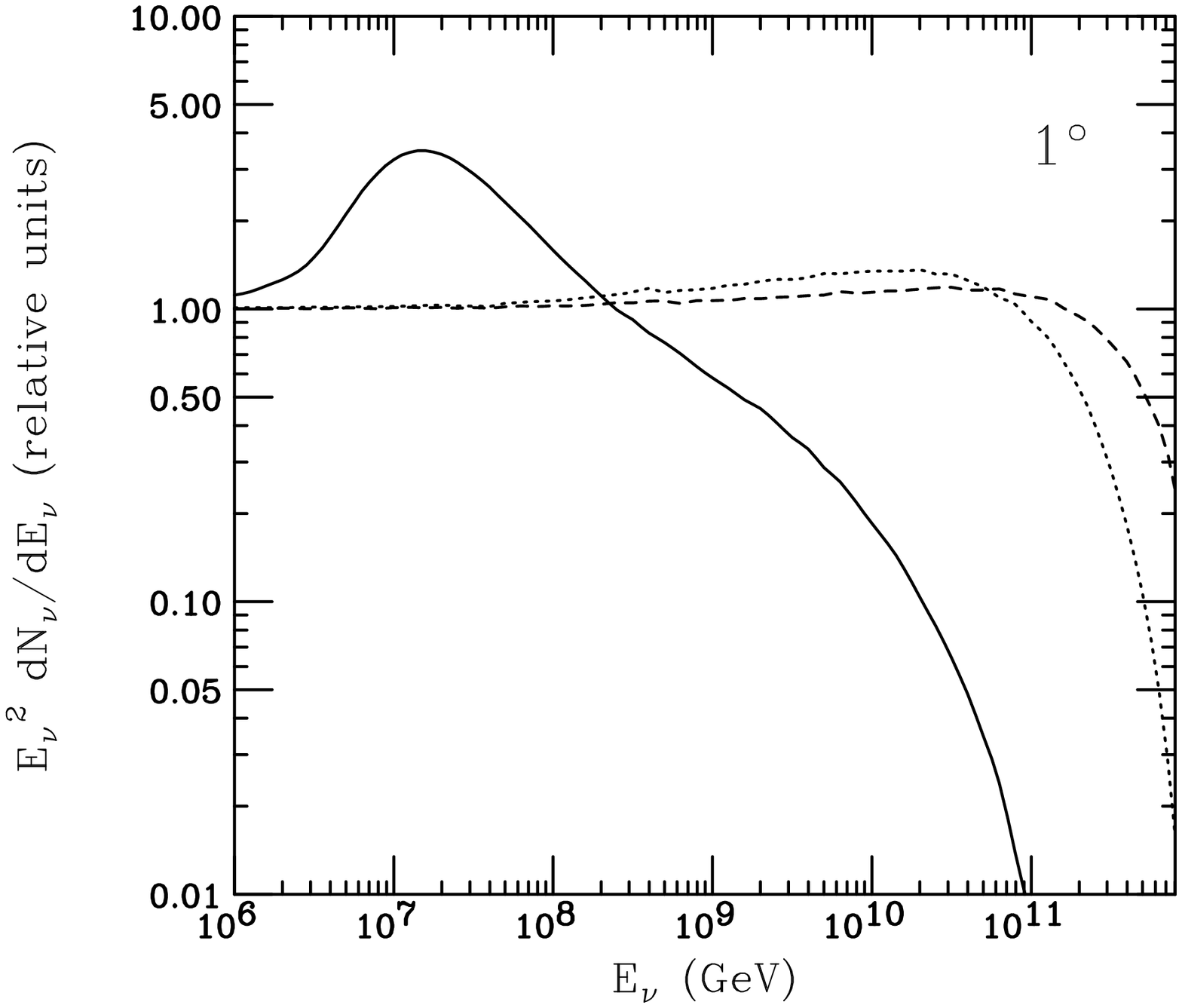} \,\,
\includegraphics[width=1.7in,angle=90]{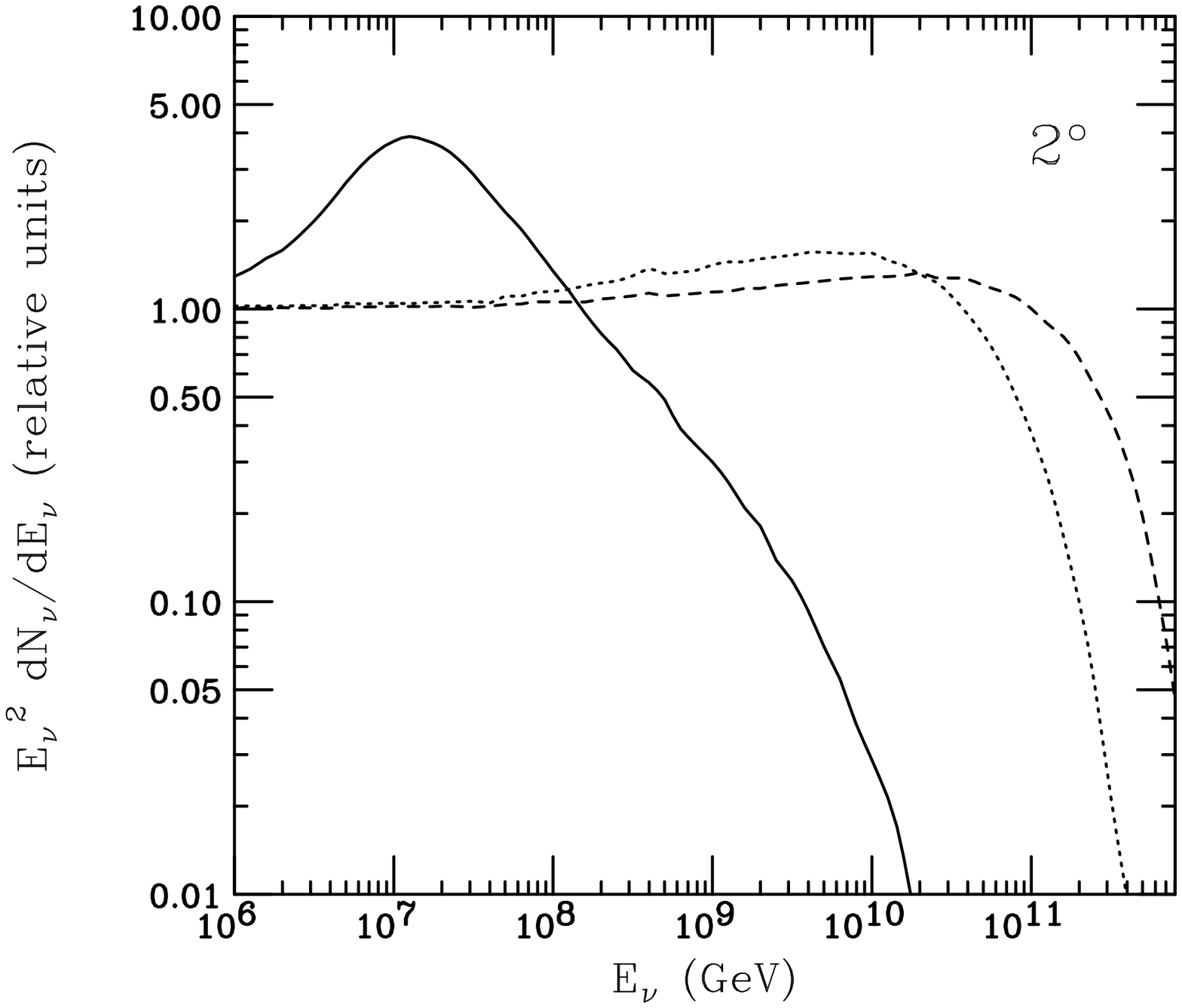} \,\,
\includegraphics[width=1.7in,angle=90]{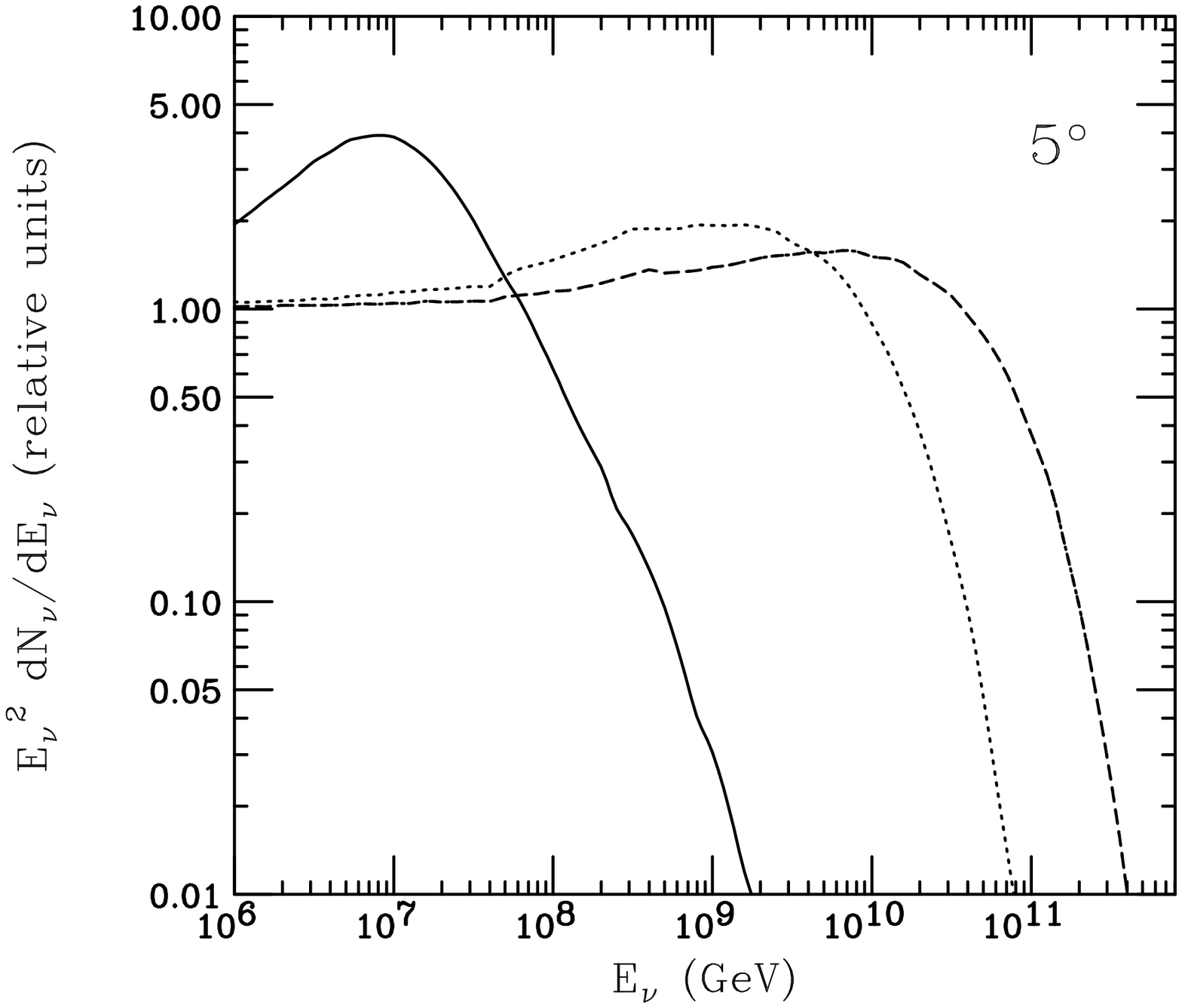}
}
\caption{The effect of interactions in the Earth on the tau neutrino
spectrum when the (CC + NC) interaction cross-section is {\em
suppressed}. As in Fig.~\ref{tauprop}, we adopt a spectrum $\propto
E^{-2}_{\nu},$ which extends to $10^{12}$ GeV. In each frame, results
are shown assuming the SM cross-section (solid-line) and a
cross-section smaller by a factor of 2 (dotted-line) and a factor of 5
(dashed-line). The three frames are for incoming angles of
1$^{\circ}$, 2$^{\circ}$ and 5$^{\circ}$ degrees below the horizon.}
\label{ox}
\end{figure}

In order to probe deviations from the (unscreened) parton model
calculation of the cross-section, it is necessary to note that the
screening corrections affect CC and NC equally. To assess the
experimental sensitivity to such effects, we assume a uniform
suppression of the cross-section by a factor of 2 or 5 and show in
Fig.~\ref{ox} the effect on the spectrum of Earth skimmers, as a
function of the incoming angle.  For a cross section reduced by a
factor of 2, the total event rate of Earth skimmers is 1.4 yr$^{-1}$,
which is slightly {\em larger} than for the unscreened parton
model. The reduction in cross-section due to screening will be energy
dependent in general, but as shown in Table~\ref{measuret}, the effect
is mainly manifest at intermediate energies of $\sim
10^8-10^{10}$~GeV, corresponding to center-of-mass energies $\sqrt{s}
\simeq 10^4 - 10^5$~GeV; at these energies the ratio of
quasi-horizontal to Earth-skimming events is a useful diagnostic of
any suppression in the cross-section. This is in fact primarily
because the cosmogenic neutrino flux peaks at these energies,
nevertheless since this represents a reasonable lower limit to the
expected flux, this sensitivity is likely to be achieved and even
surpassed. A factor of 2 reduction in the cross-section may appear
extreme, even so it is clear that Auger can probe the behavior of
parton distribution functions (pdfs) in a kinematic region out of
reach of forseeable accelerators. This will be particularly beneficial
for callibrating different hadronic interaction models of air shower
development, which presently differ significantly in their
predictions~\cite{Anchordoqui:1998nq}.

\begin{figure}[tbh]
\centering\leavevmode
\mbox{
\includegraphics[width=1.7in,angle=90]{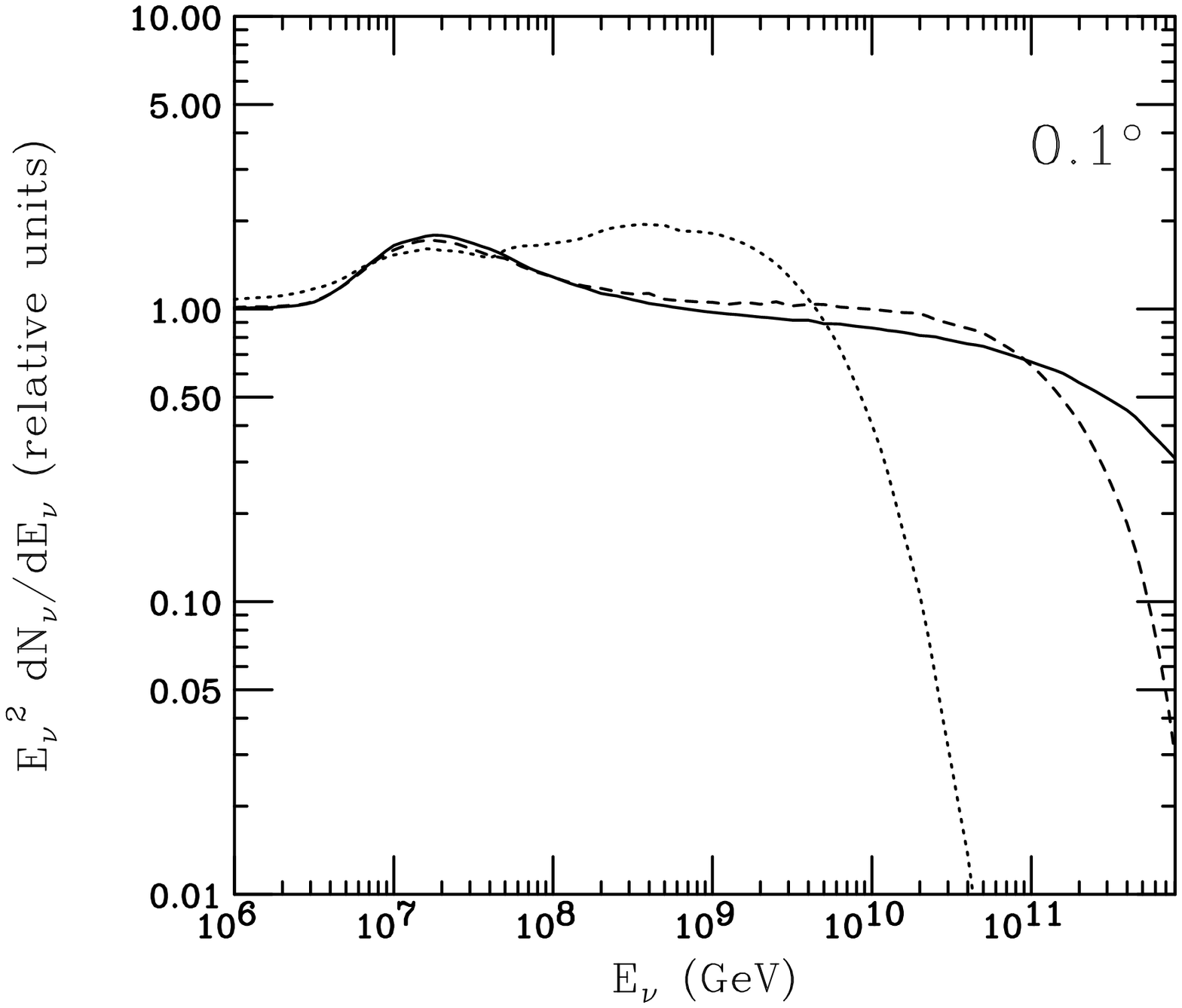} \,\,
\includegraphics[width=1.7in,angle=90]{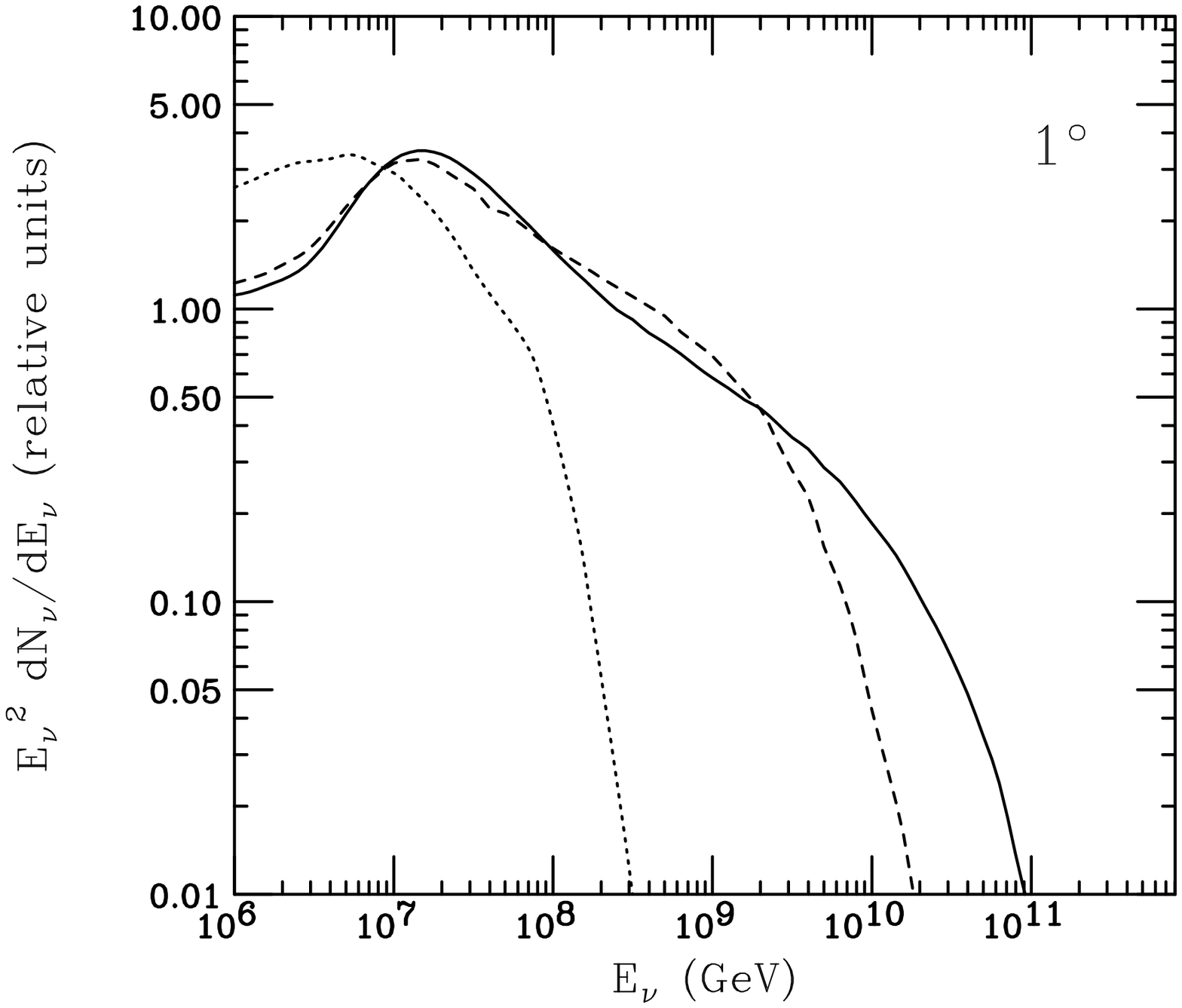} \,\,
\includegraphics[width=1.7in,angle=90]{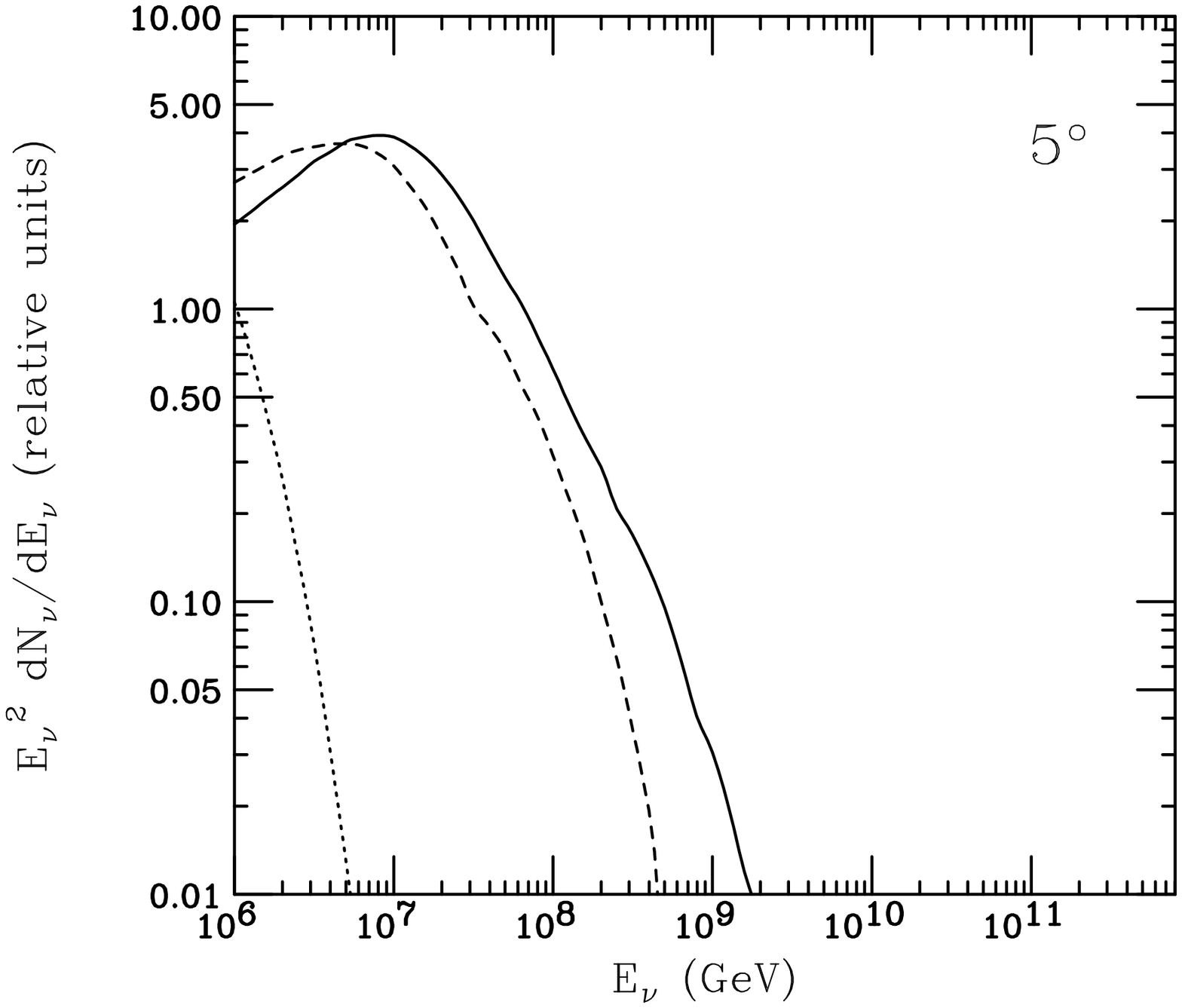}
}
\caption{The effect of interactions in the Earth on the tau neutrino
spectrum when the (NC) interaction cross-section is {\em enhanced}. As
in Fig.~\ref{tauprop}, we adopt a spectrum $\propto E^{-2}_{\nu},$
which extends to $10^{12}$ GeV. In each frame, results are shown for
the cases of the SM cross-section (solid-line), a cross-section 10
times larger (dashed-line) and 100 times larger (dotted-line). The
three frames are for incoming angles of 0.1$^{\circ}$, 1$^{\circ}$ and
5$^{\circ}$ degrees below the horizon.}
\label{absorb}
\end{figure}

With regard to enhancements of the cross-section by new physics, in
general this will be different for CC and for NC interactions. To
assess the sensitivity of Auger, we consider a toy model in which only
the NC cross-section is enhanced by a factor ranging between 3 and
100, while assuming the inelasticity to be the same as in the
SM.\footnote{This resembles the KK graviton exchange, as we discuss in the
next section.}  In Fig.~\ref{absorb} we show that this results in a
suppression of Earth-skimming tau spectrum. By contrast the
quasi-horizontal showers are enhanced, resulting in a steady increase
of the ratio of quasi-horizontals to Earth skimmers, as the NC
cross-section is increased (see Table~\ref{measurenc}). Clearly Auger
would be sensitive to substantial increases of the NC cross-section.

Thus both an increase and a decrease of the neutrino-nucleon
cross-section from the na\"{\i}ve SM value will have distinctive
observational signatures. To quantitatively assess the sensitivity of
Auger to such effects, the uncertainty in the cosmic neutrino fluxes
must also be taken into account~\cite{Anchordoqui:2005pn}. Moreover,
to determine the acceptances to different types of events, a full
detector simulation is clearly required to improve over the
approximate estimates~\cite{augersim} adopted here.

\begin{table}[!ht]
\begin{tabular} {|c| c| c| c| c| c| c|} 
\hline 
$\sigma_{\nu N}$ & $10^6 - 10^7$ & $10^7- 10^8$ & $10^8-10^9$ 
& $10^9-10^{10}$  
& $10^{10}-10^{11}$ & $10^{11}-10^{12}$ \\
\hline\hline
$\,\,\,$ \,\,\,\,\,SM $\,\,\,$& $3.6 \times 10^{-5}$ & 
0.056  & 0.85 & 0.41 & 0.020 & 
$1.1 \times 10^{-4}$ \\
\hline
$\,\,\,\textstyle{\rm SM}\, \times \frac{1}{2}\,\,$ & 
$2.1 \times 10^{-5}$ & 0.057 & 0.86 & 0.45 & 0.026 & $1.8 \times 10^{-4}$\\
\hline
\end{tabular}
\caption{Variation of the rate (in yr$^{-1}$) of Earth-skimming tau
neutrino induced events in various energy intervals (in GeV), for the
SM (unscreened parton) cross-section, and for a cross section 2 times
smaller (for {\em both} CC and NC). The cosmogenic neutrino flux has
been assumed.}
\label{measuret}
\end{table}

\begin{table}[!ht]
\begin{tabular} {|c |c c| c c |c c|} 
\hline 
$\sigma_{\nu N}$ & Quasi-horizontal & \,\, 
& Earth-skimming $\nu_{\tau}$ &\,\,& Ratio & \\
\hline\hline
Standard Model & 0.067 && 1.3 && 0.05 & \\
\hline
SM $\times$ 3 & 0.096 && 1.1 && 0.09 & \\
\hline
SM $\times$ 10 & 0.20 && 0.68 && 0.29 & \\
\hline
SM $\times$ 100 & 1.5 && 0.081 && 19 & \\
\hline
\end{tabular}
\caption{ The energy integrated rate (in yr$^{-1}$) of
quasi-horizontal and Earth-skimmers, as well as their ratio, for the
SM cross-section and for different enhancements of the NC component
alone. The cosmogenic neutrino flux has been assumed. }
\label{measurenc}
\end{table}

\subsection{Prospects for Flavor Ratio Measurements}
\label{flavormeasure}

In most models of astrophysical neutrino sources, neutrinos are
generated through the decay of charged pions: $\pi^+ \rightarrow \mu^+
\nu_{\mu} \rightarrow e^+ \nu_{e} \bar{\nu}_{\mu} \nu_{\mu}$ or $\pi^-
\rightarrow \mu^- \bar{\nu}_{\mu} \rightarrow e^- \bar{\nu}_{e}
\nu_{\mu} \bar{\nu}_{\mu}$, thus the flavor ratio at source is
$\nu_e:\nu_{\mu}:\nu_{\tau} = 1/3:2/3:0$. However, oscillations modify
this ratio as neutrinos propagate to Earth. Given the observed near
maximal mixings~\cite{Eidelman:2004wy} and the long baselines
involved, the predicted flavor ratio at Earth is
$\nu_e:\nu_{\mu}:\nu_{\tau} \approx 0.36:0.33:0.31$ following
Ref.~\cite{doublebang}.  However, cosmic (anti-)neutrinos may also be
generated in the decay of neutrons: $n \rightarrow p^+ e^-
\bar{\nu}_e$. In this case, the initial flavor ratio of
$\nu_e:\nu_{\mu}:\nu_{\tau} = 1:0:0$ becomes
$\nu_e:\nu_{\mu}:\nu_{\tau} \approx 0.56:0.26:0.18$ at
Earth~\cite{Anchordoqui:2003vc}. In either case a measured deviation
from these predictions could indicate new physics if the neutrino
production mechanism is well understood.

\begin{figure}[tbh]
\centering\leavevmode
\mbox{
\includegraphics[width=3.5in,angle=90]{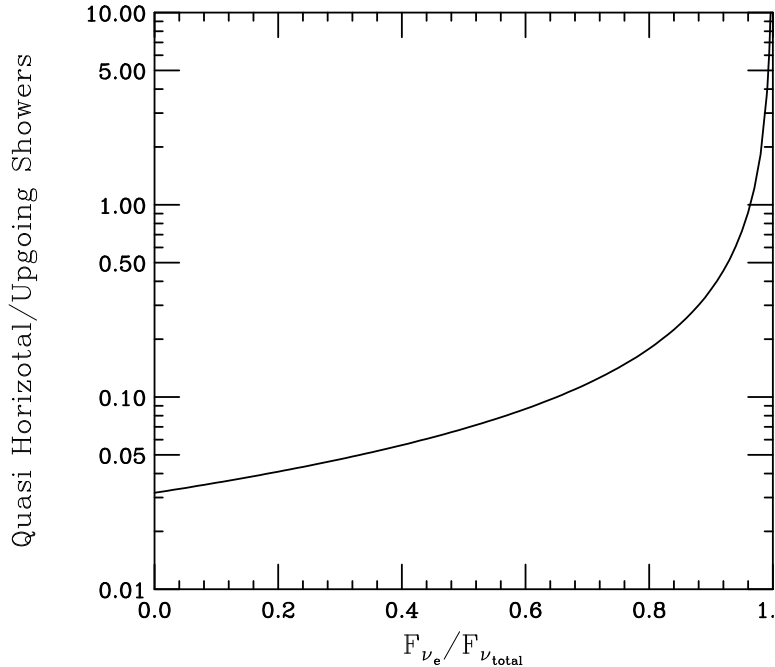}
}
\caption{The ratio of neutrino induced quasi-horizontal showers to
Earth-skimming tau neutrino induced upgoing showers as a function of
the flavor content of the cosmic neutrino flux. We have assumed equal
numbers of muon and tau neutrinos and adopted the cosmogenic neutrino
flux.}
\label{ratiomeasure}
\end{figure}

To study the sensitivity of Auger to the flavor content, we plot in
Fig.~\ref{ratiomeasure} the ratio of quasi-horizontal showers to
Earth-skimming events as the $\nu_e$ flux is varied in ratio to the
other flavors. As in Table~\ref{measurenc}, this ratio is 0.05 when
the flux is equally spread among flavors.

\section{Models of New Physics}
\label{exotic}

\subsection{Low Scale Quantum Gravity}

Two of the most important scales in physics are the Planck scale
($M_{\rm Pl} = G_{\rm N}^{-1/2} \simeq 10^{19}$~GeV) and the weak
scale ($M_W = G_{\rm F}^{-1/2} \simeq 300$~GeV), and a long standing
problem is explaining the hierarchy between these scales.  The
traditional view is to adopt $M_{\rm Pl}$ as {\em the} fundamental
scale and attempt to derive $M_W$ through some dynamical mechanism
(e.g. renormalization group evolution).  However recently several
models~\cite{Arkani-Hamed:1998rs,Randall:1999ee} have been proposed
where $M_W$ is instead the fundamental scale of nature. In the
simplest construction of these models, the SM fields are confined to a
3+1-dimensional `brane-world' (corresponding to our observed
universe), while gravity propagates in a higher dimensional `bulk'
space.

If space-time is assumed to be a direct product of a 3+1-dimensional
manifold and a flat spatial $n$-dimensional torus $T^{n}$ (of common
linear size $2\pi r_{\rm c}$), one obtains a definite representation
of this picture in which the effective 4-dimensional Planck scale is
related to the fundamental scale of gravity, $M_D$, according
to~\cite{Arkani-Hamed:1998rs}
\begin{equation}
 M_{\rm Pl}^2 = 8 \pi \,r_{\rm c}^n M_D^{n+2} \,\,,
\end{equation} 
where $D = 4 + n$. If $M_D$ is to be not much higher than the
electroweak scale, then this requires $r_{\rm c}$ to be large in
Planck units and thus reformulates the hierarchy problem.

For illustrative purposes in what follows we will consider only the
case of flat extra-dimensions. The consequences of more exoteric
scenarios (such as warped extra-dimensions~\cite{Randall:1999ee}) have
been studied in detail by various authors~\cite{Davoudiasl:1999jd}.

\subsubsection{Sub-Planckian Regime}

From our 4-dimensional point of view, the higher dimensional massless
gravitons then appear as an infinite tower of KK modes, of which the
lowest is the massless graviton itself, while its excitations are
massive. The mass-squared of each KK graviton mode reads, $m^2 =
\sum_{i=1}^n \,\ell_i^2 / r_{\rm c}^{2},$ where the mode numbers
$\ell_i$ are integers.  Note that the weakness of the gravitational
interaction is compensated by the large number of KK modes that are
exchanged: the coupling $M^{-2}_{\rm Pl}$ of the graviton vertex is
cancelled exactly by the large multiplicity of KK excitations $\sim
\hat s^{n/2} \,r_{\rm c}^n$, so that the final product is $\sim \hat
s^{n/2}/M_D^{2+n}$~\cite{Giudice:1998ck}. Here $\sqrt{\hat s}$ is the
center-of-mass energy available for graviton-KK emission. Taking brane
fluctuations into account, a form factor $\sim {\rm e}^{-m^2/M_D^2}$
is introduced at each graviton vertex~\cite{Bando:1999di}.  This
exponential suppression, which parametrizes the effects of a finite
brane tension, provides a dynamical cutoff in the (otherwise
divergent) sum over all KK contributions to a given scattering
amplitude. Altogether, one may wonder whether the rapid growth of the
cross-section with energy in neutrino-nucleon reactions mediated by
spin 2 particles carries with it observable deviations from SM
predictions.

A simple Born approximation to the elastic neutrino-parton cross
section (which underlies the total neutrino-proton cross-section)
leads, without modification, to $\hat \sigma_{\rm el} \sim \hat
s^2$~\cite{Nussinov:1998jt,Jain:2000pu}. Unmodified, this behavior by
itself eventually violates unitarity. This may be seen either by
examining the partial waves of this amplitude, or by studying the high
energy Regge behavior of an amplitude $A_R (\hat s,\hat t) \propto
\,\hat s^{\alpha(\hat t)}$ with spin-2 Regge pole, {\it viz.,}
intercept $\alpha(0)=2$. For the latter, the elastic cross-section is
given by
\begin{equation}
  \frac{{\rm d}\hat\sigma_{\rm el}}{{\rm d}\hat t}\, 
\sim\, \frac{|A_R(\hat s, \hat t)|^2}{\hat s^2}\, \sim 
  \hat s^{2\alpha(0)-2}\,\sim \hat s^2,
\end{equation} 
whereas the total cross-section reads
\begin{equation}
\hat \sigma_{\rm tot}\, \sim \frac{\Im {\rm m} 
[A_R(\hat s,0)]}{\hat s}\,\sim \hat s^{\alpha(0)-1}\,\sim 
\hat s,
\end{equation}
so that eventually, $\hat \sigma_{\rm el}> \hat\sigma_{\rm
tot}$~\cite{Anchordoqui:2000uh}.  Eikonal unitarization schemes modify
this behaviour. Specifically, for large impact parameter, a single
Regge pole exchange amplitude yields $\hat \sigma_{\rm tot} \sim
\ln^2(\hat s/s_0)$~\cite{Kachelriess:2000cb}, an estimate which is
insensitive to the underlying theory at short distances (UV
completion). Recently, the differential cross-section for such
gravity-mediated interaction at large distances has been
calculated~\cite{Emparan:2001kf}. Because of the large cross-sections,
albeit with low inelasticity, there would be distinctive double and/or
multiple bang events~\cite{Illana:2005pu} similar to those discussed
in Sec.~\ref{QH}~\cite{Anchordoqui:2004bd}.  For small impact
parameters, it becomes difficult to respect partial wave unitarity as
corrections to the eikonal amplitude are expected to become
important. Note that graviton self interactions carry factors of $\hat
t$ associated to the vertices, and thus as $\hat t$ increases, so does
the attraction among the scattered particles. Eventually it is
expected that gravitational collapse to a black hole (BH) will take
place, absorbing the initial state in such a way that short distance
effects are screened by the appearance of a
horizon~\cite{Banks:1999gd,Feng:2001ib}.

\subsubsection{Trans-Planckian Regime}

According to Thorne's hoop conjecture~\cite{Thorne:ji}, a BH forms in
a two-particle collision when and only when the impact parameter is
smaller than the radius of a Schwarzschild BH of mass equal to
$\sqrt{\hat s} \equiv \sqrt{xs}$. The total cross-section for BH
production is then,
\begin{equation}
\hat\sigma_{\rm BH} = F(n)\,\pi r_s^2(\sqrt{\hat{s}}) \,, 
\label{hoopsigma}
\end{equation} 
proportional to the area subtended by a ``hoop'' of
radius~\cite{Myers:un}
\begin{equation}
\label{schwarz}
r_s(\sqrt{\hat s}) =
\frac{1}{M_D}
\left[ \frac{\sqrt{\hat{s}}}{M_D} \right]^{\frac{1}{1+n}}
\left[ \frac{2^n \pi^{\frac{n-3}{2}}\Gamma(\frac{n+3}{2})}{n+2}
\right]^{\frac{1}{1+n}}\,,
\end{equation}
where $F(n)$ is a form factor of order unity.  Recent work has
confirmed the validity of Eq.~(\ref{hoopsigma}) and evaluated the
dimension-dependent constant $F(n)$, analytically in four
dimensions~\cite{Eardley:2002re} and numerically in higher
dimensions~\cite{Yoshino:2002tx}. In the course of collapse, a certain
amount of energy is radiated in gravitational waves by the multipole
moments of the incoming shock waves~\cite{D'Eath:hb}, leaving a
fraction $y \equiv M_{\rm BH}/\sqrt{\hat s}$ available to be emitted
through Hawking evaporation~\cite{Hawking:1975sw}. Here, $M_{\rm BH}$
is a {\it lower bound} on the final mass of the BH and $\sqrt{\hat s}$
is the center-of-mass energy of the colliding particles, taken to be
partons. This ratio depends on the impact parameter of the collision,
as well as on the dimensionality of
space-time~\cite{Yoshino:2002br}. Of course, this calculation is
purely in the framework of classical general relativity, and is
expected to be valid only for energies far above the fundamental
Planck scale $M_D$, for which curvature is small outside the horizon
and strong quantum effects are hidden behind the horizon. Extending
this formalism to center-of-mass energies close to $M_D$ requires a
better understanding of quantum gravity.

String theory provides the best hope for understanding the regime of
strong quantum gravity, and in particular for computing cross-sections
at energies close to the Planck scale~\cite{Dimopoulos:2001qe}. In
principle embedding TeV-scale gravity models in realistic string
models might facilitate the calculation of cross-sections for BHs (and
string excitations) having masses comparable to $M_D$. To be specific
we will consider embedding of a 10-dimensional low-energy scale
gravity scenario within the context of SO(32) Type I superstring
theory, where gauge and charged SM fields can be identified with open
strings localized on a 3-brane and the gravitational sector consists
of closed strings that propagate freely in the internal dimensions of
the universe~\cite{Antoniadis:1998ig}. After compactification on $T^6$
down to four dimensions, $M_{\rm Pl}$ is related to the string scale,
$M_{\rm s}$, and the string coupling constant, $g_{\rm s}$, by $M_{\rm
Pl}^2 = (2 \pi \,r_{\rm c})^6 \,M_{\rm s}^8/g_{\rm s}^2$ (hereafter,
$D=10$, i.e. $n=6$).

Subsequent to formation, the BH proceeds to
decay~\cite{Chamblin:2003wg}.  The decay of an excited spinning BH
state proceeds through several stages. The initial configuration
looses hair associated with multipole moments in a balding phase by
emission of classical gravitational and gauge radiation. Gauge charges
inherited from the initial state partons are discharged by Schwinger
emission. After this transient phase, the subsequent spinning BH
evaporates by semi-classical Hawking radiation in two phases: a brief
spin-down phase in which angular momentum is
shed~\cite{Frolov:2002xf}, and a longer Schwarzschild phase.  In the
latter the emission rate per degree of particle freedom $i$ of
particles of spin $s$ with initial total energy between $(Q, Q + {\rm
d}Q)$ is found to be~\cite{Han:2002yy}
\begin{equation}
\frac{{\rm d}\dot{N}_i}{{\rm d}Q} = \frac{\sigma_s (Q, r_s)\,\,
\Omega_{d-3}}{(d-2)\,(2\pi)^{d-1}}\,\,Q^{d-2} \left[
\exp \left( \frac{Q}{T_{\rm BH}} \right) - (-1)^{2s} \right]^{-1} \,\,,
\label{rate}
\end{equation}
where $T_{\rm BH} = 7/(4\,\pi\,r_s)$ is the BH temperature,
\begin{equation}
\Omega_{d-3} = \frac{2\,\pi^{(d-2)/2}}{\Gamma[(d-2)/2]}
\end{equation}
is the volume of a unit $(d-3)$-sphere, and $\sigma_s (Q, r_s)$ is the
absorption coefficient (a.k.a. the greybody factor). Recall that SM
fields live on a 3-brane ($d=4$), while gravitons inhabit the entire
spacetime ($d=10$). The prevalent energies of the decay quanta are of
${\cal O}(T_{\rm BH}) \sim 1/r_s$, resulting in s-wave dominance of
the final state. Indeed, as the total angular momentum number of the
emitted field increases, $\sigma_s (Q,r_s)$ rapidly gets
suppressed~\cite{Kanti:2002nr}. In the low energy limit, $Q \, r_s \ll
1,$ higher-order terms are suppressed by a factor of $3 (Q\,r_s)^{-2}$
for fermions and by a factor of $25 (Q\,r_s)^{-2}$ for gauge
bosons. For an average particle energy $\langle Q \rangle$ of ${\cal
O}(r_s^{-1})$, higher partial waves also get suppressed, although by a
smaller factor. This strongly suggests that the BH is sensitive only
to the radial coordinate and does not make use of the extra angular
modes available in the internal space~\cite{Emparan:2000rs}. A recent
numerical study~\cite{Harris:2003eg} has explicitly shown that the
emission of scalar modes into the bulk is largely suppressed with
respect to the brane emission. In order to contravene the argument of
Emparan--Horowitz--Myers~\cite{Emparan:2000rs}, the bulk emission of
gravitons would need to exhibit the opposite behavior -- a substantial
enhancement into bulk modes. There is no {\it a priori} reason to
suspect this qualitative difference between $s=0$ and $s=2$, and hence
no reason to support arguments~\cite{Cavaglia:2003hg} favoring
deviation from the dominance of visible decay. With this in mind, we
assume the evaporation process to be dominated by the large number of
SM brane modes. The lower bound on the mass radiated in the
Schwarzschild phase could be somewhat reduced at large
$b$~\cite{Yoshino:2005hi} compared to the estimate in
Ref.~\cite{Yoshino:2002br} used here. On the other hand, the effective
range of $b$ at which there is trapping is somewhat
increase~\cite{Yoshino:2005hi}, with the result that there is not any
significant change.

The total number of particles emitted is approximately equal to the BH
entropy,
\begin{equation}
S_{\rm BH} = \frac{\pi}{2}\,M_{\rm BH}\,r_s.
\end{equation}
At a given time, the rate of decrease in the BH mass is just
the total power radiated
\begin{equation}
\frac{{\rm d}\dot{M}_{\rm BH}}{{\rm d}Q} = - \sum_{i} c_i\, 
\frac{\sigma_s(Q, r_s)}{8 \,\pi^2}\,\,Q^3 \left[
\exp \left( \frac{Q}{T_{\rm BH}} \right) - (-1)^{2s} \right]^{-1}\,\, ,
\label{rate2}
\end{equation}
where $c_i$ is the number of internal degrees of freedom of particle
species $i$. Integration of Eq.~(\ref{rate2}) leads to
\begin{equation}
\dot{M}_{\rm BH} = - \sum_i c_i \,\,f\,\, \frac{\Gamma_s}{8\,\pi^2} \,\, \,
\Gamma(4) \,\,
\zeta(4)\, \,T^4_{\rm BH}\,A_4,
\label{m}
\end{equation}
where $f=1$ (7/8) for bosons (fermions), and the greybody factor was
conveniently written as a dimensionless constant, $\Gamma_s =
\sigma_s(\langle Q \rangle,r_s)/A_4$, normalized to the BH surface
area~\cite{Emparan:2000rs}
\begin{equation}
A_4 = \frac{36}{7}\,\pi\,\left( \frac{9}{2} \right)^{2/7}\ \, r_s^2
\label{area}
\end{equation}
seen by the SM fields ($\Gamma_{s=1/2} \approx 0.33$ and $\Gamma_{s=1}
\approx 0.34$~\cite{greybody}). Now, since the ratio of degrees of
freedom for gauge bosons, quarks and leptons is 29:72:18 (excluding
the Higgs boson), from Eq.~(\ref{m}) one obtains a rough estimate of
the mean lifetime,
\begin{equation}
\tau_{_{\rm BH}} \approx  1.67 \times 10^{-27}\,{\rm s}\, 
\left(\frac{M_{\rm BH}}{M_{10}}\right)^{9/7}
\left(\frac{\rm TeV}{M_{10}}\right) \,,
\label{lifetime}
\end{equation}
which indicates that BHs evaporate near-instantaneously into visible
quanta.

The semi-classical description outlined above is reliable only when
the energy of the emitted particle is small compared to the BH mass,
i.e.
\begin{equation}
T_{\rm BH} \ll M_{\rm BH}\,, \,\,\, 
{\rm or}\,\,\, {\rm equivalently,} \,\,\, M_{\rm BH} \gg M_{10} \,,
\label{condition}
\end{equation}
because it is only under this condition that both the gravitational
field of the brane and the back reaction of the metric during the
emission process can safely be neglected~\cite{Preskill:1991tb}. For
BHs with initial masses well above $M_{10}$, most of the decay process
can be well described within the semi-classical
approximation. However, the condition stated in Eq.~(\ref{condition})
inevitably breaks down during the last stages of evaporation. At this
point it becomes necessary to introduce quantum considerations. To
this end we turn to a quantum statistical description of highly
excited strings.

It is well-known that the density of string states with mass between
$M$ and $M+{\rm d}M$ cannot increase any faster than $\rho (M) = {\rm
e}^{\beta_{\rm H} M}/M,$ because the partition function,
\begin{equation}
Z (\beta) = \int_0^\infty {\rm d}M\, \rho(M) \,\,{\rm e}^{-M\, \beta} \,\,,
\end{equation}
would fail to converge~\cite{Hagedorn:st}. Indeed, the partition
function converges only if the temperature is less than the Hagedorn
temperature, $\beta_{\rm H}^{-1}$, which is expected to be $\sim
M_{\rm{s}}$. As $\beta$ decreases to the transition point $\beta_{\rm
H}$, the heat capacity rises to infinity because the energy goes into
the many new available modes rather than into raising the kinetic
energy of the existing particles~\cite{Frautschi:1971ij}. In the
limit, the total probability diverges, indicating that the canonical
ensemble is inadequate for the treatment of the system. However, one
can still employ a microcanonical ensemble of a large number of
similar isolated systems, each with a given fixed energy $E$. With the
center-of-mass at rest, $E = M$ so the density of states is just
$\rho(M)$ and the entropy $S = \ln \rho(M)$. In this picture,
equilibrium among systems is determined by the equality of the
temperatures, defined for each system as
\begin{equation}
T \equiv \left(\frac{\partial S}{ \partial M} \right)^{-1} = 
\frac{M}{\beta_{\rm H} M -1} \ .
\end{equation}
Equilibrium is achieved at maximum entropy when the total system heat
capacity, $C$, is positive.  Ordinary systems (on which our intuition
is founded) have $C>0$. However, for a gas of massive superstring
excitations the heat capacity,
\begin{equation}
C \equiv -\frac{1}{T^2} \left(\frac{\partial^2 S}{\partial M^2}\right)^{-1} 
= - \left(\frac{M}{T}\right)^2\, ,
\end{equation}
is {\em negative}, as is the case for BHs~\cite{Hawking:de}. The
positivity requirement on the total specific heat implies that strings
and BHs cannot coexist in thermal equilibrium, because any subsystem
of this system has negative specific heat, and thus the system as a
whole is thermodynamically unstable. This observation suggests that
BHs may end their Hawking evaporation process by making a transition
to an excited string state with higher entropy, avoiding the singular
zero-mass limit~\cite{Bowick:1985af}. The suggestion of a string
$\rightleftharpoons$ BH transition is further strengthened by three
other facts: (i) in string theory, the fundamental string length
should set the minimum value for the Schwarzschild radius of any
BH~\cite{Veneziano:1986zf}; (ii) $T_{\rm BH} \sim \beta_{\rm H}^{-1}$
for $r_s \sim M_{\rm s}^{-1}$~\cite{Susskind:ws}; (iii) there is an
apparent correlation between the greybody factors in BH decay and the
level structure of excited strings~\cite{Das:1996wn}. The string
$\rightleftharpoons$ BH ``correspondence
principle''~\cite{Horowitz:1996nw} unifies these concepts: When the
size of the BH horizon drops below the size of the fundamental string
length $\ell_{\rm s} \gg \ell_{10},$ where $\ell_{10}$ is the
fundamental Planck length, an adiabatic transition occurs to an
excited string state. Subsequently, the string will slowly lose mass
by radiating massless particles with a nearly thermal spectrum at the
unchanging Hagedorn temperature~\cite{Amati:1999fv}.(Note that the
probability of a BH radiating a large string, or of a large string
undergoing a fluctuation to become a BH is negligibly
small~\cite{Horowitz:1997jc}.)

The continuity of the cross-section at the correspondence point, at
least parametrically in energy and string coupling, provides an
independent supporting argument for this
picture~\cite{Dimopoulos:2001qe}.  Specifically, in the perturbative
regime, the Virasoro-Shapiro amplitude leads to a ``string ball'' (SB)
production cross-section $\propto g_{\rm s}^2 \hat{s}/M_{\rm
s}^4$. This cross-section saturates the unitarity bounds at $g_{\rm
s}^2 \hat{s}/M_{\rm s}^2 \sim 1$~\cite{Amati:1987wq}, so before
matching the geometric BH cross-section $\propto r_s^2$, there is a
transition region at which $\hat{\sigma} \sim M_{\rm s}^{-2}$. All in
all, the rise with energy of the parton-parton $\rightarrow$ SB/BH
cross-section can be parametrized as~\cite{Dimopoulos:2001qe}
\begin{eqnarray}
\hat \sigma (\sqrt{\hat{s}})\sim\left\{
\begin{array}{ll}
\displaystyle
\frac{g_{\rm s}^2\,\hat{s}}{M_{\rm s}^4} &\qquad M_{\rm s} \ll 
\sqrt{\hat{s}} \leq M_{\rm s}/ g_{\rm s}\,,\\
\displaystyle
\frac{1}{M_{\rm s}^2}&\qquad M_{\rm s}/ g_{\rm s} < \sqrt{\hat{s}} 
\leq M_{\rm s}/ g_{\rm s}^2\,,\\
\displaystyle
\frac{1}{M_{10}^2}\,\left[\frac{\sqrt{\hat{s}}}{M_{10}}\right]^{2/7}
&\qquad M_{\rm s}/ g_{\rm s}^2<\sqrt{\hat{s}}\,,
\end{array}
\right.
\end{eqnarray}
where $M_{10} = (8\pi^5)^{1/8}\,M_{\rm s}/g_{\rm s}^{1/4}$

\begin{figure}[tbh]
\centering\leavevmode
\mbox{
\includegraphics[width=3.5in,angle=0]{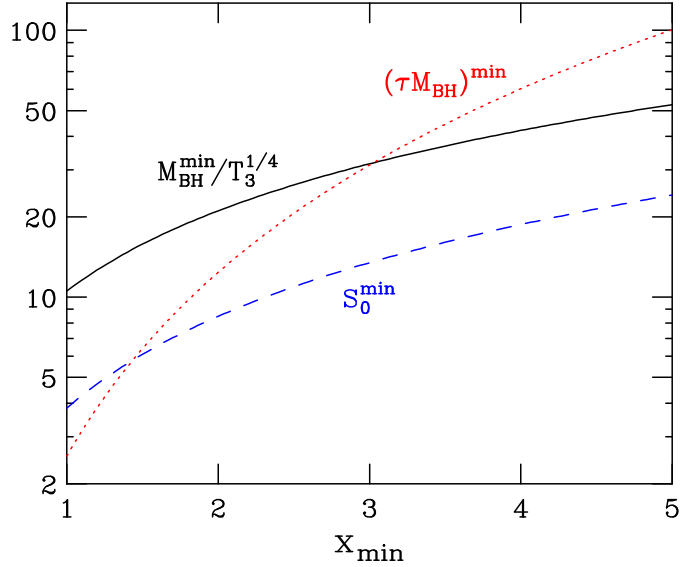}
}
\caption{Quantitative measures of the validity of the semi-classical
analysis of BH production for $n=6$ extra dimensions, where $x_{\rm
min} \equiv M_{\rm BH, min}/ M_{10}$.}
\label{fig:xminplot}
\end{figure}

\begin{figure}[tbh]
\centering\leavevmode
\mbox{
\includegraphics[width=3.5in,angle=0]{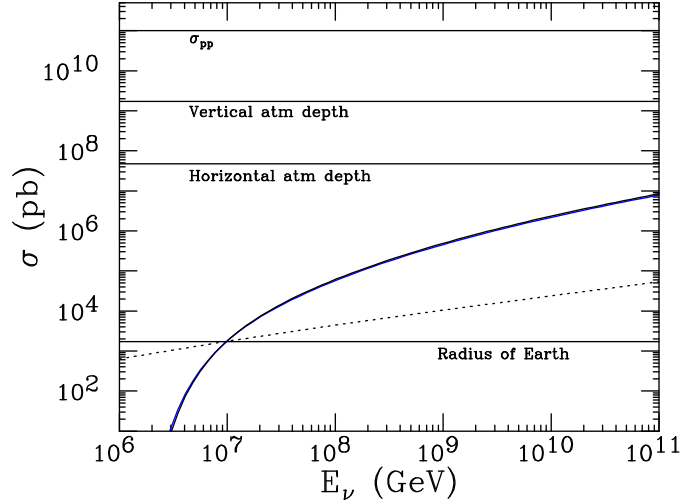}
}
\caption{The cross-section for BH production in neutrino nucleon
collisions, for $n = 6$ extra dimensions, assuming $M_{10} = 1$~TeV
and $M_{\rm BH, min} = M_{10}.$ Energy losses by gravitational
radiation have been included. The SM $\nu N$ cross-section is
indicated by the dotted line. For comparison the typical $pp$
cross-section is shown, as well as the cross-section required for
triggering vertical and horizontal atmospheric showers. The
cross-section for absorption by the Earth is also
shown~\cite{Anchordoqui:2004xb}.}
\label{bhcross}
\end{figure}

The inclusive production of BHs proceeds through different final
states for different classical impact parameters
$b$~\cite{Yoshino:2002br}. These final states are characterized by the
fraction $y(z)$ of the initial parton center-of-mass energy,
$\sqrt{\hat s}=\sqrt{xs}$, which is trapped within the horizon. Here,
$z= b/b_{\rm max},$ where $b_{\rm max}= 1.3 \, r_s(\sqrt{\hat
s})$~\cite{Yoshino:2002br}. With a lower cutoff $M_{\rm BH,min}$ on
the BH mass required for the validity of the semi-classical
description, this implies the joint constraint
\begin{equation}
 y(z)\,\,\sqrt{x s} \ge M_{\rm BH,min}
\label{constraint}
\end{equation}
on the parameters $x$ and $z$. Because of the monotonically decreasing
nature of $y(z)$, Eq.~(\ref{constraint}) sets an {\it upper} bound
$\bar z(x)$ on the impact parameter for fixed $x.$ The corresponding
parton-parton BH cross-section is $\hat \sigma_{_{\rm BH}} (x) = \pi
\bar b^2(x),$ where $\bar b=\bar z b_{\rm max}.$ The total BH
production cross-section is then~\cite{Anchordoqui:2003jr}
\begin{equation}
\sigma_{_{\rm BH}}(E_\nu,M_{\rm BH,min},M_{10}) \equiv 
\int_{\frac{M_{\rm BH,min}^2}{
y^2(0) s}}^1 \, dx
\,\sum_i f_i(x,Q) \ \hat \sigma_{_{\rm BH}}(x) \,\,,
\label{sigma}
\end{equation}
where $i$ labels parton species and the $f_i(x,Q)$ are
pdfs~\cite{Pumplin:2002vw}.  The momentum scale $Q$ is taken as
$r_s^{-1},$ which is a typical momentum transfer during the
gravitational collapse process. The parameter $M_{\rm BH, min}$ plays
in important role n interpreting the results derived below. The
validity of the semi-classical calculation requires at least three
criteria to be satisfied.  First, $S_0$, the initial entropy of the
produced BH should be large enough to ensure a well-defined
thermodynamic description~\cite{Preskill:1991tb}. Second, the BH
lifetime $\tau_{\rm BH}$ should be large compared to its inverse mass
so that the black hole behaves like a well-defined resonance. Third,
the BH mass must be large compared to the scale of the 3-brane tension
$T_3$ so that the brane does not significantly perturb the BH
metric~\cite{Frolov:2002as}. Quantitative measures of these three
criteria are given in Fig.~\ref{fig:xminplot} for $n=6$, assuming $T_3
= \sqrt{8\pi}/(2\pi)^6 \ M_{10}^4$ for 6 toroidally-compactified
dimensions~\cite{Polchinski:1996na}. We see that all three criteria
are adequately satisfied for $M_{\rm BH,\ min} \agt 3
M_{10}$~\cite{Anchordoqui:2003ug}. The resulting $\nu N \to$ BH
production cross-section is shown in Fig.~\ref{bhcross}.

\begin{figure}[tbh]
\centering\leavevmode
\mbox{
\includegraphics[width=3.5in,angle=90]{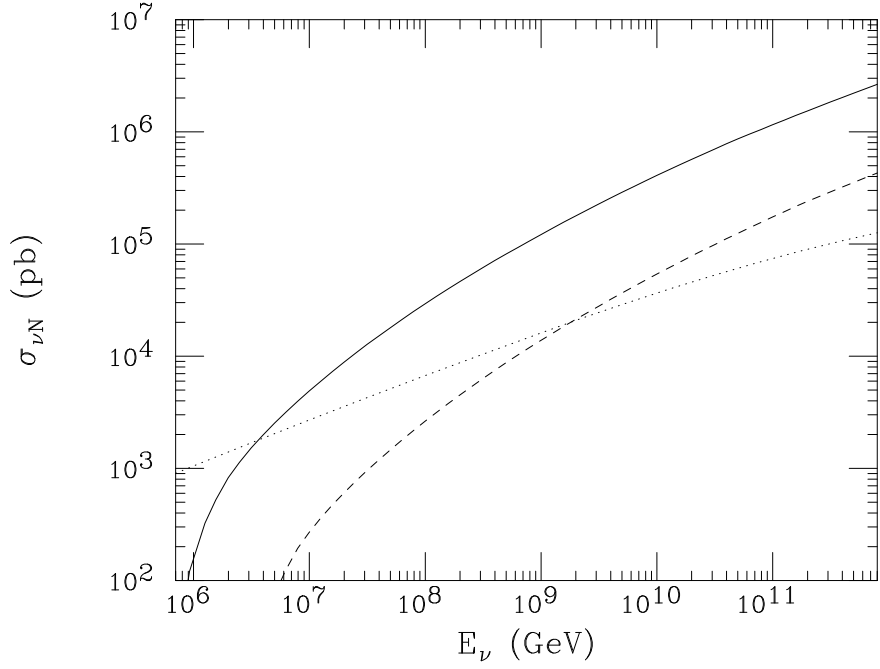}
}
\caption{The neutrino-nucleon cross-section including the effects of
TeV scale string resonances. The solid and dashed lines correspond to
models with string tension $M_{\rm s}$ = 1 and 2 TeV,
respectively. The Standard Model cross-section is shown as a dotted
line for comparison.}
\label{stringsigma}
\end{figure}

In the perturbative string regime, i.e. $M_{\rm SB,min} <
\sqrt{\hat{s}} \leq M_s/g_s$, the SB production cross-section is taken
to be 
\begin{equation}
\sigma_{_{\rm SB}}(E_\nu,M_{\rm SB,min}, M_{10}) = 
\int_{\frac{M_{\rm SB,min}^2}{s}}^1
{\rm d}x   \,\sum_i f_i(x,Q) \, \hat\sigma_{_{\rm SB}}(\hat{s}) \,,
\end{equation}
where $\hat\sigma_{_{\rm SB}} (\hat{s})$ contains the Chan-Paton
factors which control the projection of the initial state onto the
string spectrum.  In general, this projection is not uniquely
determined by the low-lying particle spectrum, so there are one or
more arbitrary constants.  The analysis in the $\nu q \rightarrow \nu
q$ channel illustrates this point~\cite{Cornet:2001gy}. The $\nu g$
scattering, relevant for $\nu N$ interactions at ultra-high energies,
introduces additional ambiguities. In our calculations we adopt the
estimates given in Ref.~\cite{stringy2} considering the saturation
limit and including both neutrino-quark and neutrino-gluon
scattering. The resulting $\nu N \to$ SB cross-section is shown in
Fig.~\ref{stringsigma}, setting the Chan-Paton factors equal to 1/2.
 
\begin{figure}[tbh]
\centering\leavevmode
\mbox{
\includegraphics[width=2.6in,angle=90]{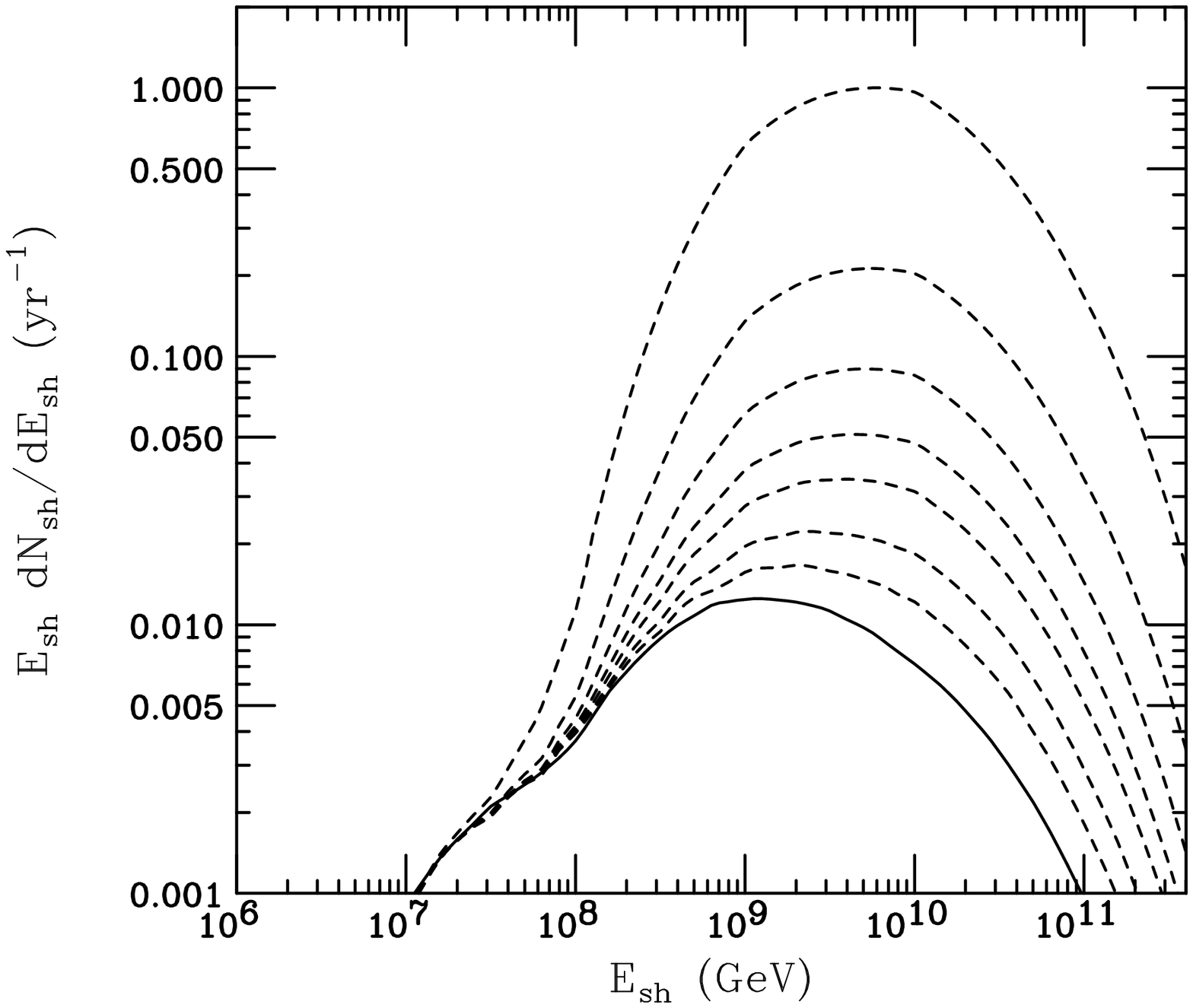}
\includegraphics[width=2.6in,angle=90]{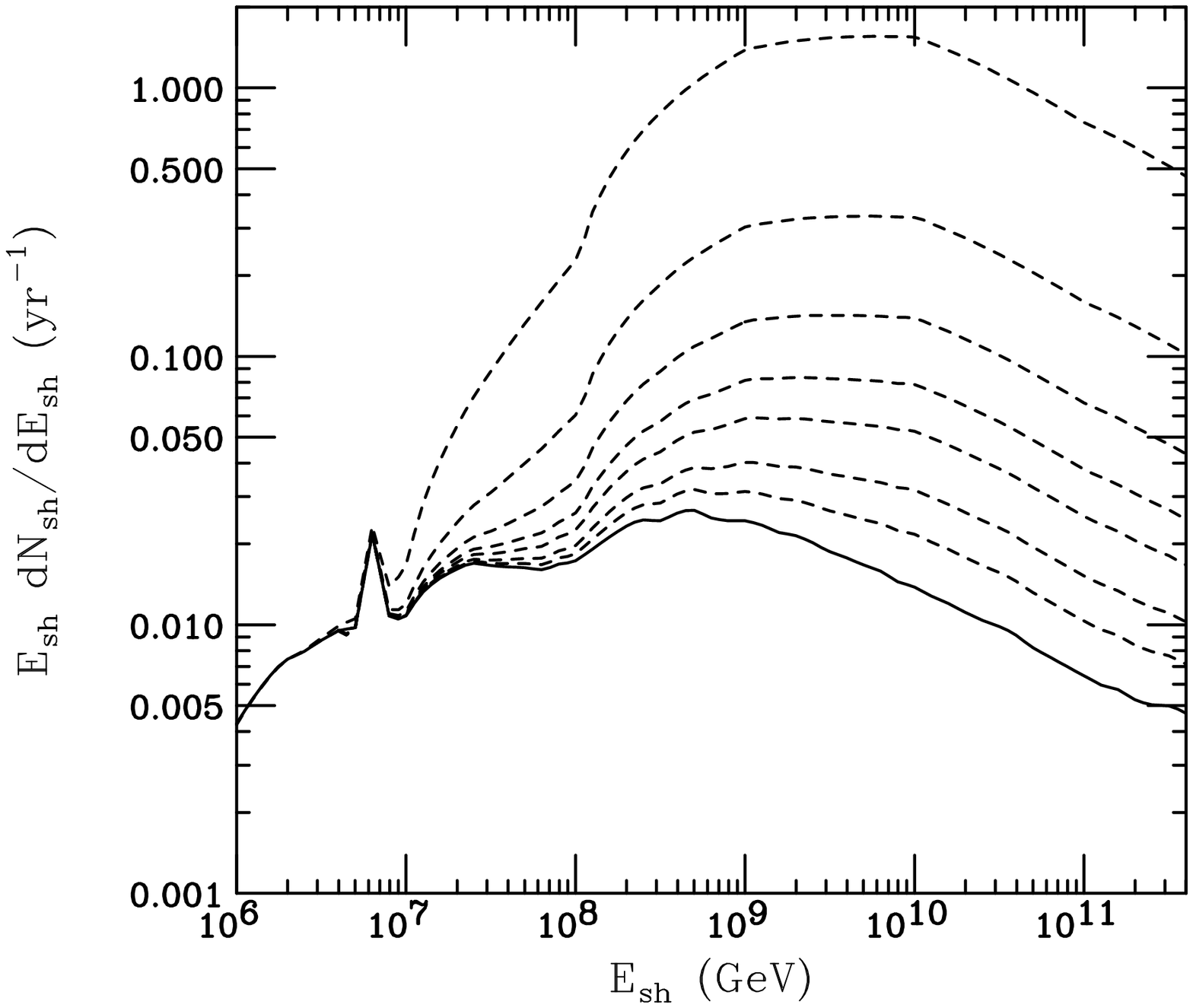}}
\caption{The spectrum of quasi-horizontal, deeply penetrating, black
hole induced showers as would be seen by Auger for the cosmogenic flux
(left) and the Waxman-Bahcall flux (right). The dashed
lines indicates different values of the fundamental Planck scale (from
below $M_{10} = \,10,\, 7, \,5, \,4, \,3, \,2, \,1$~TeV; in all cases
$M_{\rm BH,min} = 3 M_{10}$) while the solid line is the SM
prediction.}
\label{qhbh}
\end{figure}

\begin{figure}[tbh]
\centering\leavevmode
\mbox{
\includegraphics[width=2.6in,angle=90]{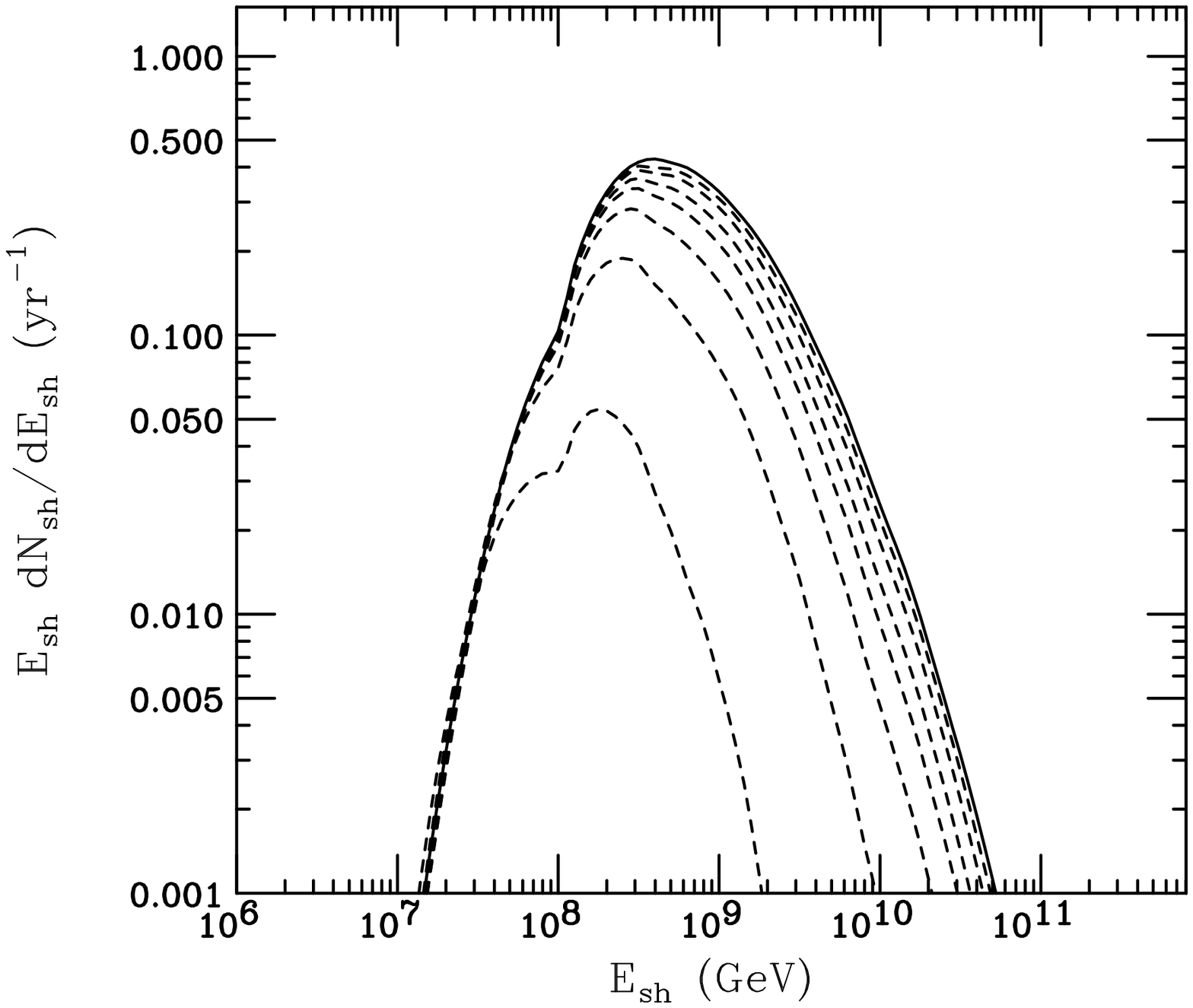}
\includegraphics[width=2.6in,angle=90]{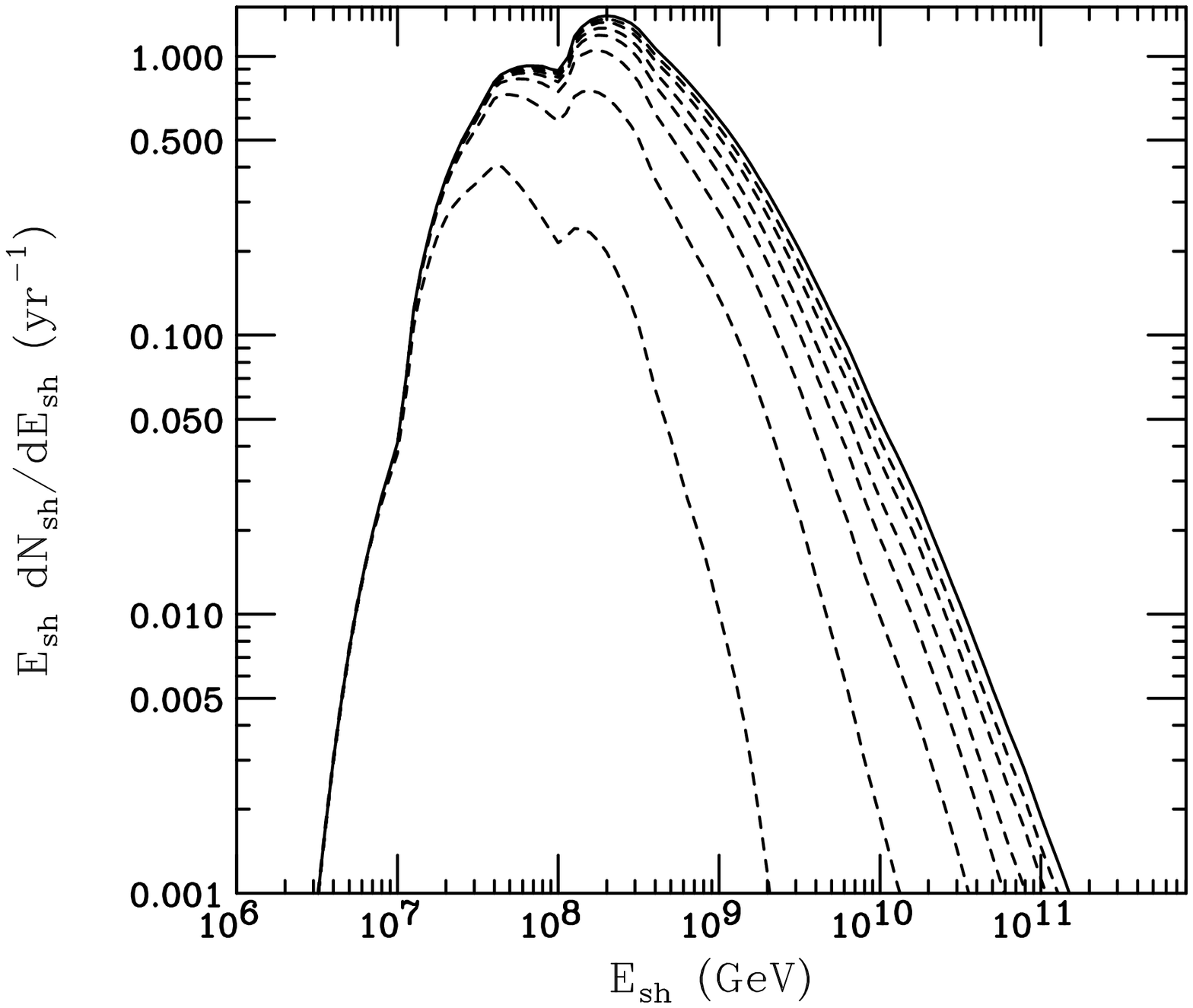}
}
\caption{The spectrum of Earth skimming, tau neutrino black hole
induced showers as would be seen by Auger for the cosmogenic flux
(left) and the Waxman-Bahcall flux (right). The dashed
lines indicates different values of the fundamental Planck scale (from
below $M_{10} = \,1,\, 2, \,3, \,4, \,5, \,7, \,10$~TeV; in all cases
$M_{\rm BH, min} = 3 M_{10}$), while the solid line is the SM
prediction.}
\label{taubh}
\end{figure}

As can be seen in Figs.~\ref{bhcross} and \ref{stringsigma}, although
the neutrino interaction length is reduced below the SM value due to
BH/SB production, it is still far larger than the Earth's atmospheric
depth. Neutrinos therefore would produce BH/SBs with roughly equal
probability at any point in the atmosphere. As a result, the light
descendants of the BH/SB may initiate low-altitude, quasi-horizontal
showers at rates significantly higher than SM
predictions.\footnote{Additionally, neutrinos that traverse the
atmosphere unscathed may produce black holes via interactions in
the ice or water and be detected by neutrino
telescopes~\cite{Kowalski:2002gb}.} Because of this the atmosphere
provides a buffer against contamination by hadronic showers (for which
the electromagnetic component is completely attenuated at such large
zenith angles) allowing a good characterization of BH-induced showers
when $S \gg 1$~\cite{Feng:2001ib,Dutta:2002ca}.

\begin{figure}[tbh]
\centering\leavevmode
\mbox{
\includegraphics[width=2.6in,angle=90]{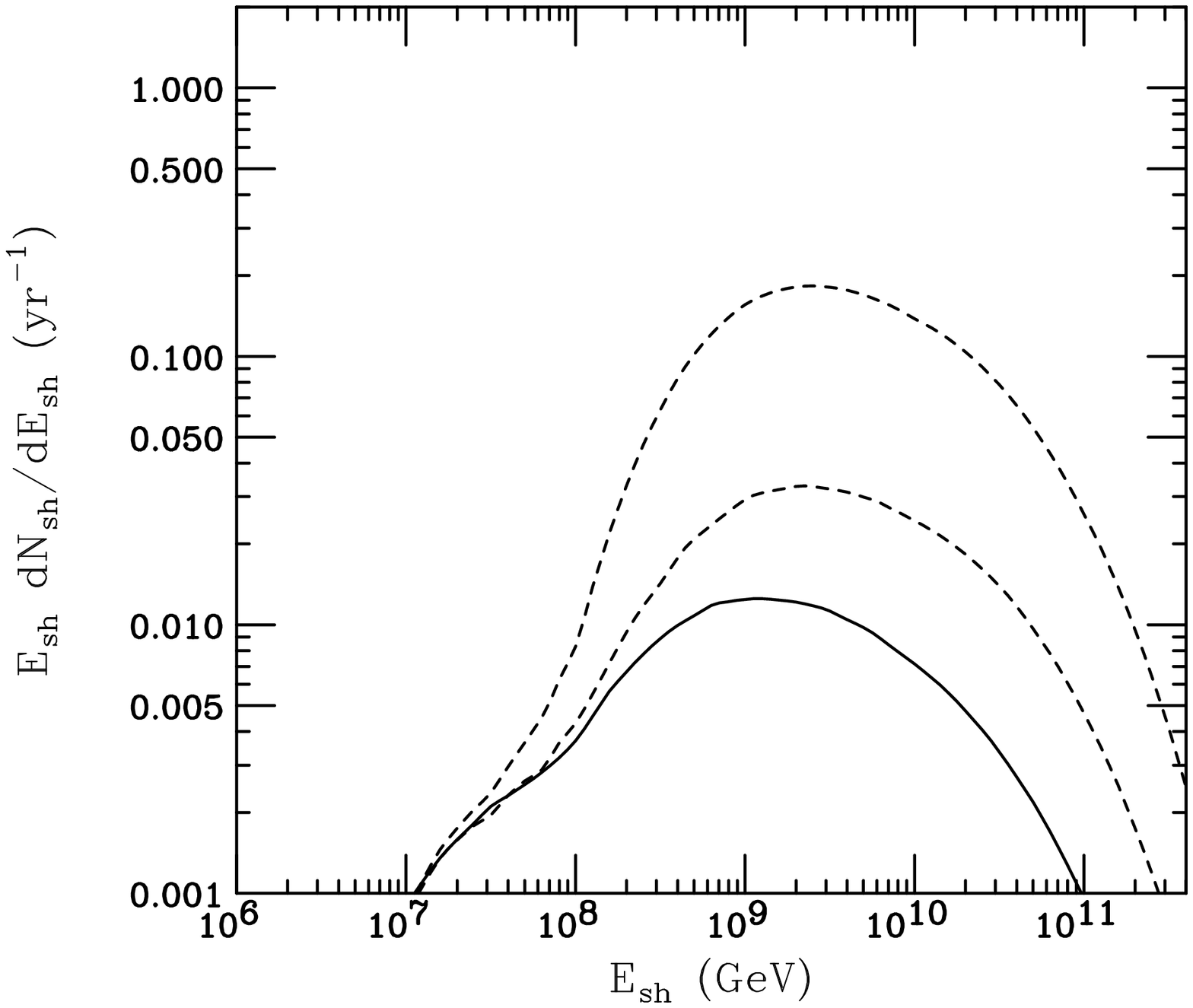}
\includegraphics[width=2.6in,angle=90]{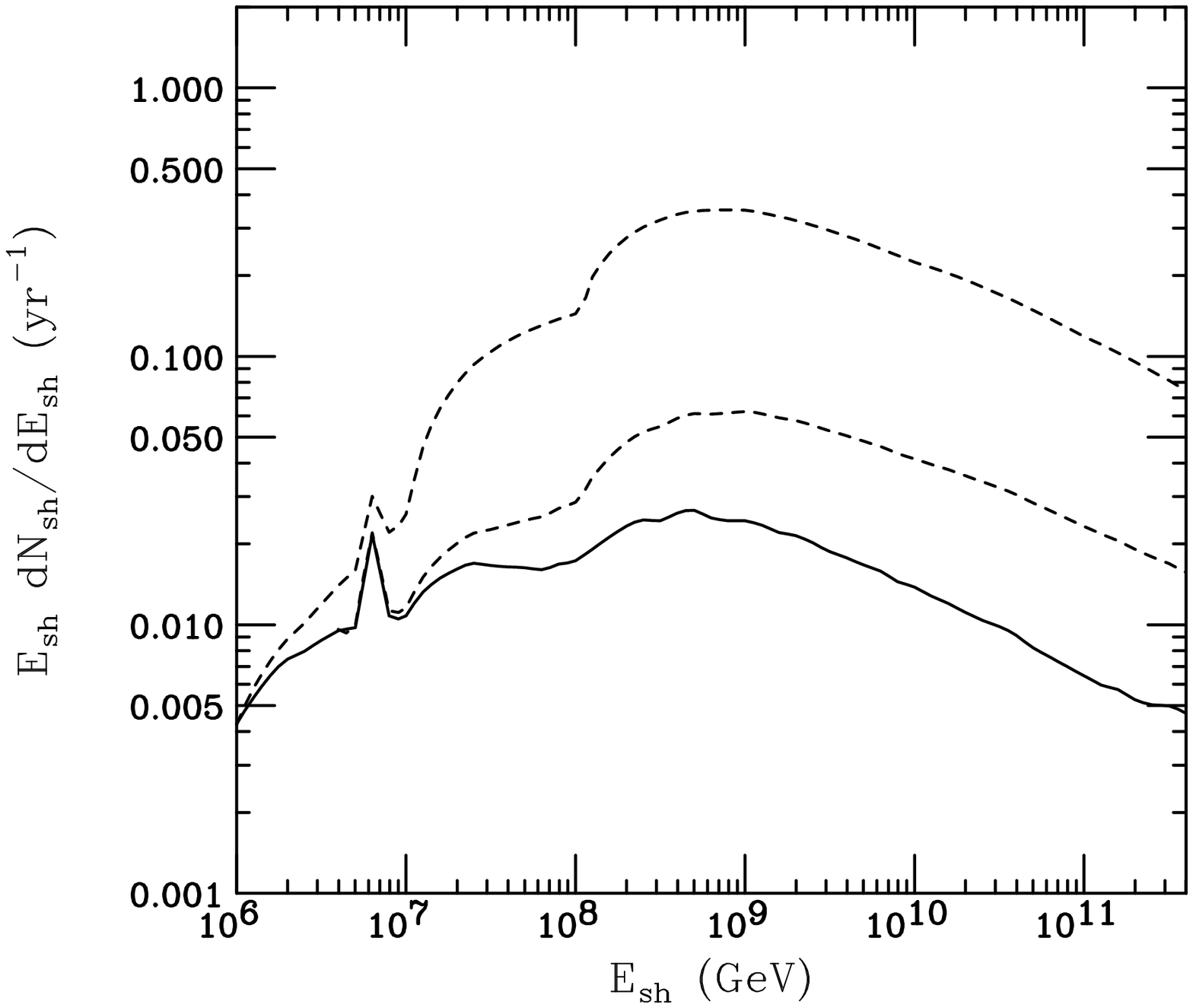}
}
\caption{The spectrum of quasi-horizontal, deeply penetrating,
neutrino string-ball induced showers as would be seen by Auger for the
cosmogenic flux (left) and the Waxman-Bahcall flux (right). The dashed
lines refer to the string scale $M_{\rm s} = 1$ TeV (upper) and
$M_{\rm s} = 2$ TeV (lower), while the solid line is the SM
prediction.}
\label{qhstring}
\end{figure}

\begin{figure}[tbh]
\centering\leavevmode
\mbox{
\includegraphics[width=2.6in,angle=90]{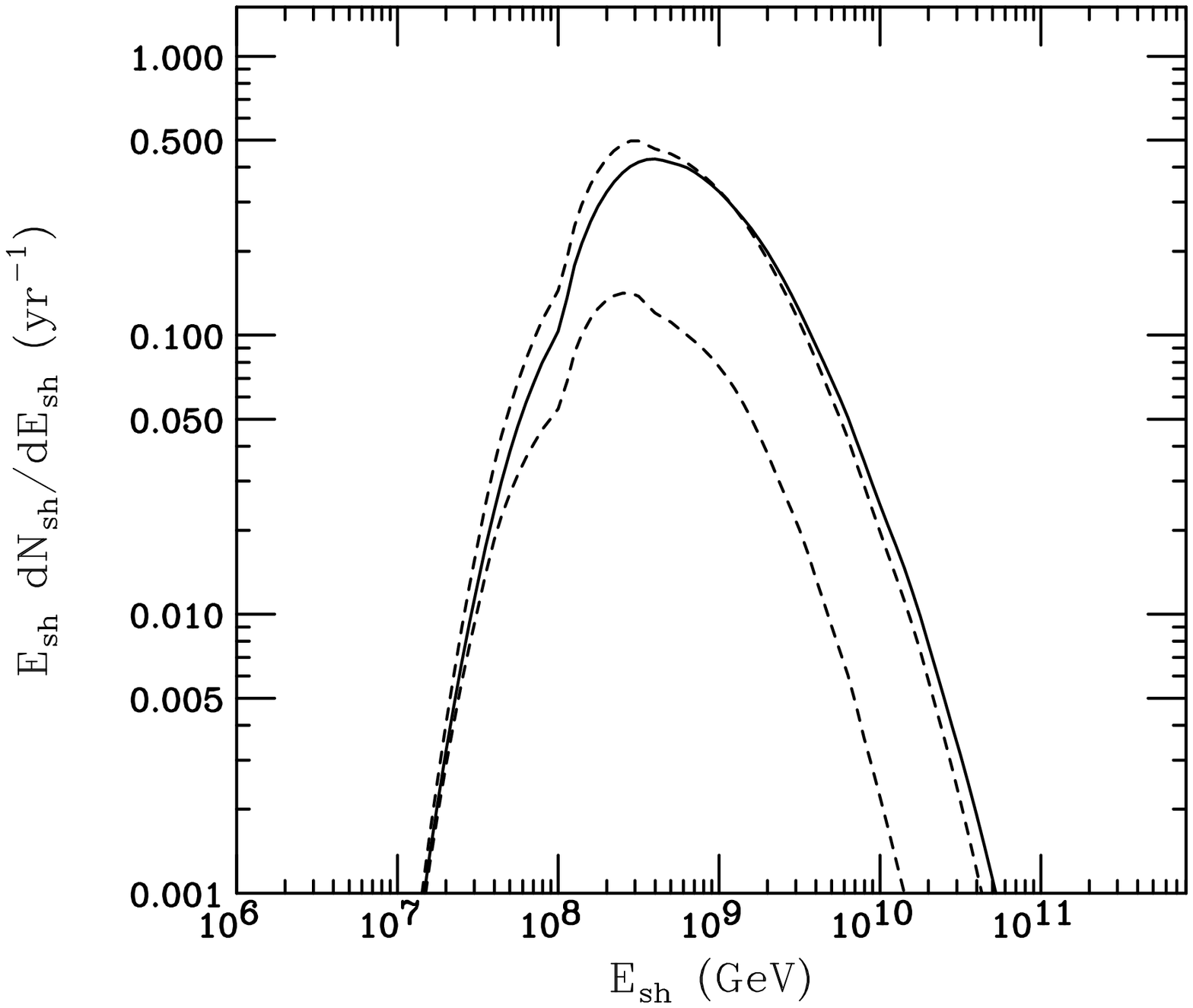}
\includegraphics[width=2.6in,angle=90]{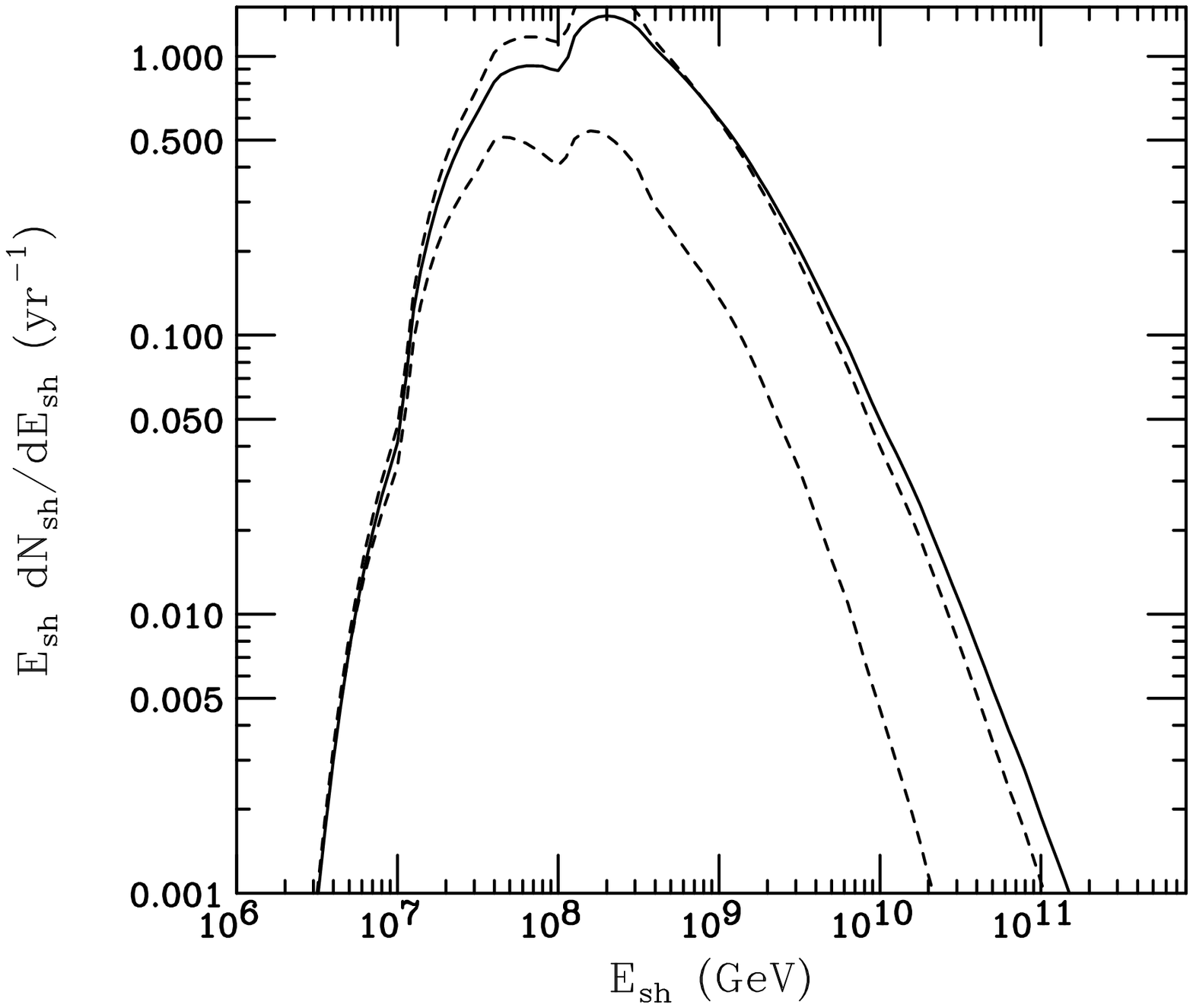}
}
\caption{The spectrum of Earth skimming, tau neutrino string ball
induced showers as would be seen by Auger for the cosmogenic flux
(left) and the Waxman-Bahcall flux (right). The dashed lines refer to
the string scale $M_{\rm s} = 1$ TeV (upper) and $M_{\rm s} = 2$ TeV
(lower), while the solid line is the SM prediction.}
\label{taustring}
\end{figure}

If the quasi-horizontal deep shower rate is found to be anomalously
large, it can be ascribed either to an enhancement of the incoming
neutrino flux, or to an enhancement in the neutrino-nucleon
cross-section. However, these possibilities may be distinguished by
focusing on events which arrive at very small angles to the
horizon. An enhanced flux will increase {\em both} the
quasi-horizontal and Earth-skimming event rates, whereas a large BH
cross-section {\em suppresses} the latter, because the hadronic decay
products of BH evaporation do not escape the Earth's
crust~\cite{Anchordoqui:2001cg}. To quantify the potential of Auger in
discriminating BH/SB induced showers, we show separately the BH
production event rates for quasi-horizontal and Earth skimming
neutrinos in Figs.~\ref{qhbh} and \ref{taubh}. The SB production rates
are similarly given in Figs.~\ref{qhstring} and \ref{taustring} and a
summary of these event rates is provided in Tables~\ref{bhtable} and
\ref{stringtable} respectively.

\begin{table}[!ht] 
\begin{tabular}{|c|| c| c|| c| c|| c| c|} 
\hline
\multicolumn{1}{|c||}{$\sigma_{\nu N}$} & 
\multicolumn{2}{c||}{Quasi-horizontal} &
\multicolumn{2}{c||}{Earth-skimming $\nu_{\tau}$} &
\multicolumn{2}{c|}{Ratio}\\ 
\hline 
& Cosmogenic & Waxman-Bahcall & Cosmogenic & Waxman-Bahcall & Cosmo & WB\\
\hline\hline 
Standard Model  & 0.067 & 0.22 & 1.3  & 5.0 & 0.050 & 0.044 \\
\hline
$M_{10}=$ 1 TeV & 4.4   & 10.6 & 0.13 & 1.0 & 36    & 10.2  \\ 
\hline 
$M_{10}=$ 2 TeV & 0.95  & 2.4  & 0.48 & 2.6 & 2.0   & 0.91  \\
\hline 
$M_{10}=$ 3 TeV & 0.42  & 1.1  & 0.77 & 3.5 & 0.54  & 0.3   \\ 
\hline 
$M_{10}=$ 4 TeV & 0.25  & 0.66 & 0.96 & 4.1 & 0.26  & 0.16  \\
\hline 
$M_{10}=$ 5 TeV & 0.18  & 0.48 & 1.1  & 4.4 & 0.16  & 0.11  \\ 
\hline 
$M_{10}=$ 7 TeV & 0.12  & 0.34 & 1.2  & 4.7 & 0.1   & 0.073 \\ 
\hline
$M_{10}=$ 10 TeV& 0.089 & 0.27 & 1.3  & 4.8 & 0.08  & 0.056 \\ 
\hline
\end{tabular} 
\caption{Black hole producing event rates of quasi-horizontal showers
and Earth-skimming tau neutrino induced showers expected to be
observed per year by Auger for both the cosmogenic neutrino flux and
the Waxman-Bahcall flux. In all cases $M_{\rm BH, min} = 3 M_{10}$.}
\label{bhtable}
\end{table}

\begin{table}[!ht]
\begin{tabular}{|c|| c| c|| c| c|| c| c|} 
\hline
\multicolumn{1}{|c||}{$\sigma_{\nu N}$} & 
\multicolumn{2}{c||}{Quasi-horizontal} &
\multicolumn{2}{c||}{Earth-skimming $\nu_{\tau}$} &
\multicolumn{2}{c|}{Ratio}\\ 
\hline 
& Cosmogenic & Waxman-Bahcall & Cosmogenic & Waxman-Bahcall & Cosmo & WB\\
\hline \hline
Standard Model & 0.067 & 0.22 & 1.3 & 5.0 & 0.05 & 0.044 \\
\hline
$M_s=$ 1 TeV & 0.86 & 2.5 & 0.4 & 2.0 & 2.1 & 1.3 \\
\hline
$M_s=$ 2 TeV & 0.17 & 0.48 & 1.5 & 5.7 & 0.11 & 0.084 \\
\hline
\end{tabular}
\caption{String ball producing event rates of quasi-horizontal
showers and Earth-skimming tau neutrino induced showers expected to be
observed per year by Auger for both the cosmogenic neutrino flux and
the Waxman-Bahcall flux.}
\label{stringtable}
\end{table}

\subsection{Non-perturbative Electroweak Interactions}

The transition probability between two flat space vacua can be
calculated in a Minkowski framework in analogy with WKB tunneling
through non-vacuum fluctuations, or by evaluating the minimal action
appropriate to a classical solution of Euclidean space in a given
topological sector~\cite{Belavin:1975fg}. As is well known, in Yang-Mills
theories the inclusion of massless fermions fundamentally alters the
picture~\cite{'tHooft:1976fv}: transitions between vacua (separated by
energy barriers whose minimum height is set by the sphaleron energy
$E_{\rm sp}$~\cite{Klinkhamer:1984di}) will be totally suppressed
unless accompanied by the simultaneous emission or absorption of {\em
all} fermions coupled to the gauge field.  In the Minkowski
description, these fermions emerge during level-shifting in the strong
${\cal O}(1/g)$ gauge fields interpolating between vacua ($g=$
coupling constant). In the Euclidean description, the
presence of a zero mode $\omega$ for each light fermion coupled to the
gauge field will, because of the rules of Grassman integration,
generate a 't Hooft vertex~\cite{'tHooft:1976fv} with all the
different fermions appearing as legs,
\begin{equation}
{\cal L}_{\rm eff} \propto  \prod_{i=1\dots N} \overline \omega
F_i + {\rm h.c.} \; ,
\end{equation}
where $F_i$ is a chiral fermion field. Whether these exotic processes
occur with sizeable rates in high energy particle collisions is a
long-standing open question~\cite{Aoyama:1986ej}.

At center-of-mass energies $\sqrt{\hat s} < E_{\rm sp} \approx \pi
M_W/\alpha_{W} \approx 7.5~{\rm TeV}$, the cross-section for
electroweak instanton mediated processes is known to have an
exponential form~\cite{McLerran:1989ab}.  Here, $m_W = 80.423$~GeV
is the W$^\pm$ boson mass and $\alpha_W (m_W) = 0.0338$ is the
$SU(2)$ fine structure constant~\cite{Eidelman:2004wy}. Including
essential pre-exponential factors~\cite{Khoze:1990bm}, one has, for
the phenomenologically interesting case of fermion-fermion scattering
${\rm f+f}\stackrel{I}{\to}{\rm all}$,
\begin{eqnarray}
\nonumber
\hat\sigma_{\rm ff}^{(I)} 
&\approx & \frac{1}{m_W^2}
\,
\left( \frac{2\pi}{\alpha_W}\right)^{7/2}
\,
\exp\left[ -\frac{4\pi}{\alpha_W}\,
F_{\rm hg} \left( \frac{\sqrt{\hat s}}{4\pi m_W/\alpha_W} \right)\right]
\\[1.5ex] \label{cross-qfd} & \simeq &
5.3\times 10^3\ {\rm mb}\  
\exp\left[ -\frac{4\pi}{\alpha_W}\,
F_{\rm hg} \left( \frac{\sqrt{\hat s}}{4\pi m_W/\alpha_W} \right)\right]
\,.
\end{eqnarray}
where $F_{\rm hg}$ is the ``holy-grail''
function~\cite{Mattis:1991bj}.  By means of perturbative calculations
of the relevant exclusive amplitudes about the instanton ($I$),
squaring them and summing over the final states, or, alternatively, by
means of a perturbative calculation of the forward elastic scattering
amplitude about the widely separated instanton anti-instanton
($I\overline{I}$) pair and determining the imaginary part to get the
total cross-section via the optical theorem, one may calculate the
decisive tunneling suppression exponent $F_{\rm hg}$, as a series in
fractional powers of $\epsilon \equiv \sqrt{\hat s}/(4\pi
m_W/\alpha_W) \simeq \sqrt{\hat s}/ (30\ {\rm
  TeV})$~\cite{Khoze:1990bm},
\begin{equation}
\label{FW-pert}
F_{\rm hg} (\epsilon ) = 1 
- \frac{3^{4/3}}{2}\,  \epsilon^{4/3} + \frac{3}{2}\,\epsilon^2 + 
{\mathcal O}(\epsilon^{8/3})
\,.
\end{equation}
Therefore, the total cross-section given in Eq.~(\ref{cross-qfd}) is
exponentially growing for $\epsilon\ll 1$. At $\epsilon$ of ${\mathcal
O}(1)$, however, the perturbative expression in Eq.~(\ref{FW-pert}) no
longer applies. In this energy regime, only extrapolations
of, and lower bounds on, the tunneling suppression
exponent are available~\cite{Ringwald:2002sw}.

\begin{figure}[tbh]
\centering\leavevmode
\mbox{
\includegraphics[width=3.5in,angle=0]{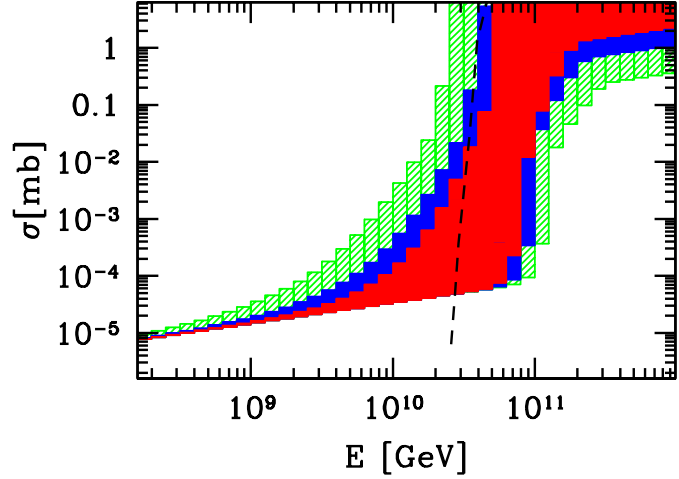}
}
\caption{The allowed 90\%, 95\%, and 99\% CL regions for interpolation
between the electroweak and QCD-like neutrino-nucleon cross-section
consistent with existing data. Also shown with a dashed line is the
predicted enhancement of the cross-section by electroweak
sphalerons. For details see Ref.~\cite{Ahlers:2005zy,Han:2003ru}.}
\label{andreas}
\end{figure}

\begin{figure}[tbh]
\centering\leavevmode
\mbox{
\includegraphics[width=2.6in,angle=90]{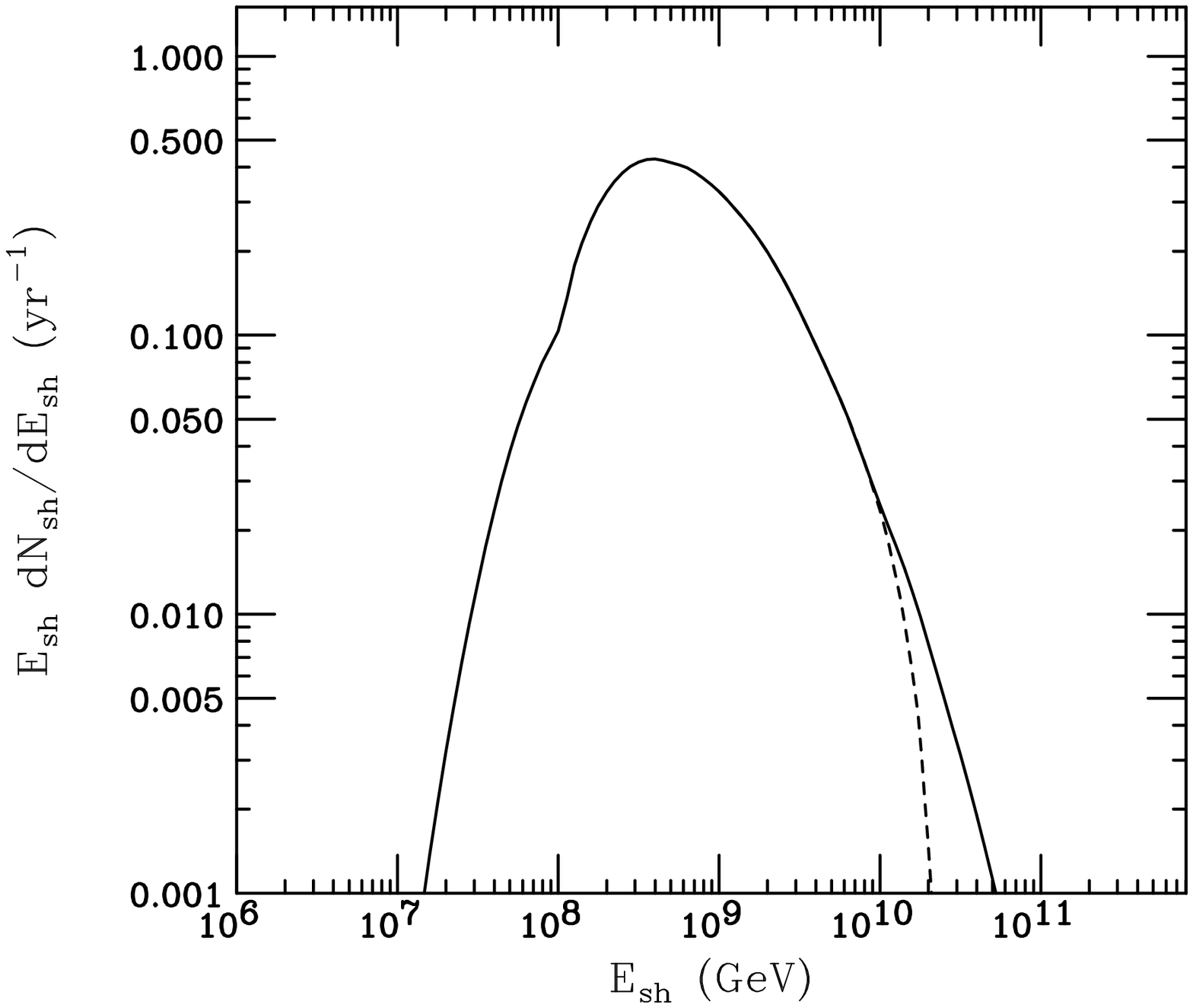}
\includegraphics[width=2.6in,angle=90]{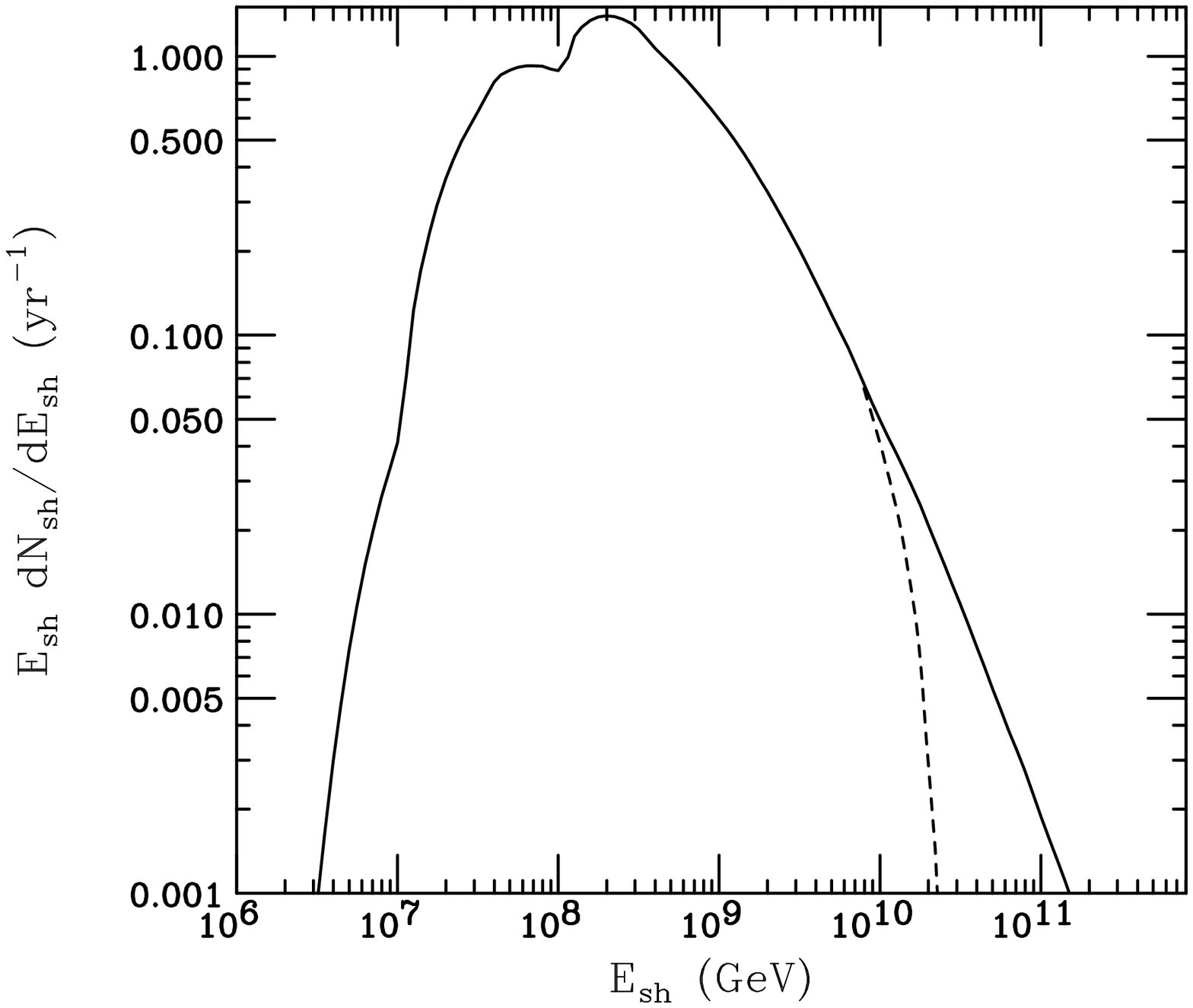}
}
\caption{Suppression of Earth skimming events due to electroweak
sphalerons as would be seen by Auger for the cosmogenic neutrino flux
(left), and for the Waxman-Bahcall flux (right). The solid line is the SM
prediction.}
\label{augerinstau}
\end{figure}

\begin{figure}[tbh]
\centering\leavevmode
\mbox{
\includegraphics[width=2.6in,angle=90]{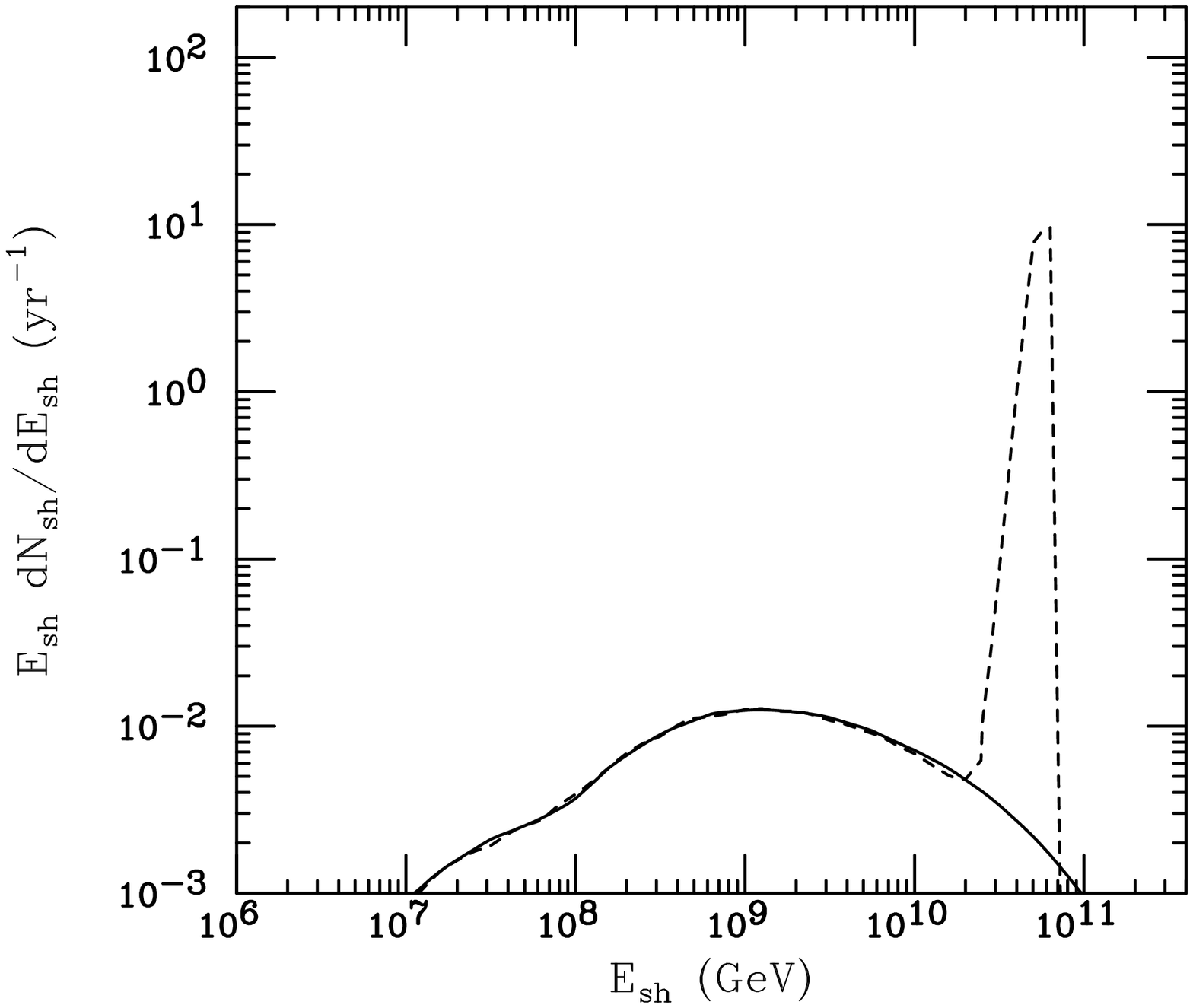}
\includegraphics[width=2.6in,angle=90]{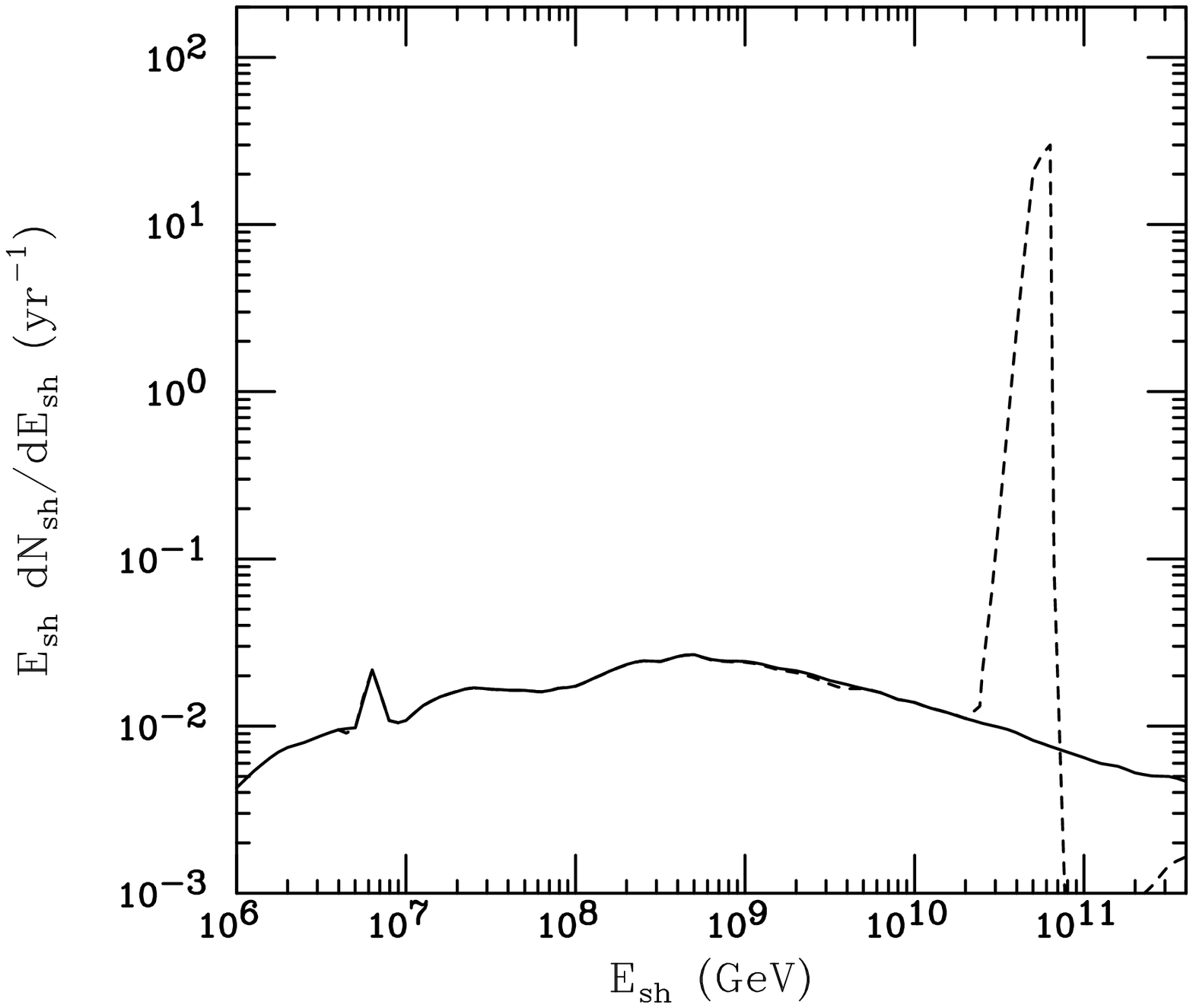}
}
\caption{The spectrum of quasi-horizontal showers mediated by
electroweak sphalerons as would be seen by Auger for the cosmogenic
neutrino flux (left), and for the Waxman-Bahcall flux (right). The
solid line is the SM prediction.}
\label{augerinsqh}
\end{figure}

Interestingly, at $\sqrt{\hat s} \sim 100$~TeV, the cross-section can
rise to values characteristic of QCD interactions. Since the
electroweak instanton-induced interaction applies equally to all
fermions, neutrinos can thus acquire hadron-like cross sections at
high energies. Moreover, the inelasticity of the process is {\em
high}.  Together, these facts imply that neutrino interactions
mediated by instantons would induce air showers in the upper
atmosphere with characteristics similar to those of proton-induced
showers~\cite{Fodor:2004tr}. Conversely, the non-observation to date
of deeply penetrating air showers constrains any sudden rise of the
neutrino-nucleon cross-section~\cite{Morris:1993wg,Ahlers:2005zy}.
Figure~\ref{andreas} shows the allowed region for transition from
electroweak to QCD-like neutrino-nucleon cross-section, consistent
with existing data. The dashed line indicates the neutrino-nucleon
cross-section taken from Ref.~\cite{Han:2003ru} obtained taking $\hat
\sigma_{ff} \agt 1~{\rm mb}$~\cite{Ringwald:2002sw}. As can be seen in
Fig.~\ref{andreas}, this prediction is marginally consistent with the
region allowed by current data. For this cross-section, the expected
event rate at Auger would be 4.3 quasi-horizontal showers per year
assuming the cosmogenic neutrino flux, and 14 quasi-horizontal showers
per year assuming the Waxman-Bahcall neutrino flux; the rate of
Earth-skimmers is 1.3 per year in both cases. As shown in
Fig.~\ref{augerinstau}, the suppression of Earth-skimmers due to
absorption in the Earth is negligible. However, the rate of
quasi-horizontal showers is increased by about 2 orders of magnitude,
and such events would be concentrated in a small energy range, as
indicated in Fig.~\ref{augerinsqh}. This would provide a {\em clean}
signal for electroweak instanton-induced interactions. Thus, if no
deeply developing showers are observed, tighter constraints can be
placed on this model, and more generally on any sudden rise in the
neutrino-nucleon cross-section.

\subsection{Neutrino Decay}

Neutrinos are known to be sufficiently light that they are stable
against tree-level electroweak decays. Moreover, decays of the form
$\nu_i \to \nu_j \gamma$ or $\nu \to \nu \nu \overline \nu$ are
severely constrained by experiment~\cite{Eidelman:2004wy}. However,
some models of lepton number violation postulate the existence of a
massless Goldstone boson, the Majoron, $X$.  Consequently, decays such
as $\nu_i \rightarrow \nu_j X$ or $\nu_i \rightarrow \overline{\nu}_j
X$, are then possible, where $\nu_{i,j}$ denote mass eigenstates
\cite{Schechter:1981cv}. Presently such possibilities are only weakly
constrained by Solar neutrino data, which set the bound $\tau/m
\gtrsim 10^{-4}$ s/eV~\cite{Bahcall:1986gq}. However, by studying
cosmic neutrinos which have travelled over far longer baselines Auger
can be more sensitive to their instability by a factor of $\sim10^2 -
10^4$, if an effective flavor ratio measurement can be made. Because
of the extremely large energies probed by Auger, it will be
complementary in this regard to the IceCube neutrino
telescope~\cite{bdecay,Anchordoqui:2005gj}.

The ratio of flavors observed in the cosmic neutrino spectrum depends
on whether any species of neutrinos have decayed and on the decay
channel. In the simple situation where all heavy neutrino species
decay into the lightest mass eigenstate (or into non-interacting
states, such as a sterile neutrino), we would expect to observe at
Earth the flavor ratio
\begin{equation}
\phi_{\nu_e}:\phi_{\nu_{\mu}}:\phi_{\nu_{\tau}} = 
\cos^2\theta_{\odot} : \frac{1}{2} \sin^2 \theta_{\odot} 
:\frac{1}{2} \sin^2 \theta_{\odot} \approx 6:1:1, 
\end{equation}
where $\theta_{\odot}$ is the solar neutrino mixing angle and we have
assumed the normal hierarchy as well as $U_{e3} =0$.  This result is
independent of the flavor ratio at source.  However, for the case of
an inverted hierarchy, the predicted flavor ratio at Earth is
\begin{equation}
\phi_{\nu_e}:\phi_{\nu_{\mu}}:\phi_{\nu_{\tau}} = 
U^2_{e3}:U^2_{\mu 3}:U^2_{\tau 3} \approx 0:1:1,
\end{equation}
where $U_{\alpha_i}$ is the neutrino mixing matrix and we have taken
the atmospheric mixing angle to be maximal.  These results are in
striking contrast to the expectation for stable neutrinos discussed
earlier in Sec.~\ref{flavormeasure}.

These two cases represent the most extreme deviations from the usual
phenomenology and are robust in that they do not depend on the flavor
composition at source. A variety of other (more baroque) possibilities
have been considered, e.g. only the heaviest neutrino eigenstate
decays, but the predicted flavor ratios after propagation then depend
on the assumed flavor ratio at source~\cite{bdecay}.  In
Table~\ref{decaytable} we list some of these possibilities assuming
the usual mass hierarchy and source flavor ratios as for pion decay.

We cannot measure the flavor ratios directly at Auger. However, as
discussed earlier, Earth-skimming events are generated uniquely by tau
neutrinos, while quasi-horizontal showers can be generated by all
neutrino flavors. Furthermore, because of maximal mixing of $\nu_\mu$
and $\nu_\tau$ we expect their fluxes to be always
comparable. Therefore, by combining these two measurements Auger can
potentially determine the flavor ratios of ultra-high energy
neutrinos.

\begin{table}[!ht]
\begin{tabular} {|c|c|c|} 
\hline
Decaying Mass Eigenstates & Decay Products & 
$\phi_{\nu_e}:\phi_{\nu_{\mu}}:\phi_{\nu_{\tau}}$ \\
\hline\hline 
$\nu_3$, $\nu_3$ & Irrelevant & 6:1:1 \\
\hline
$\nu_3$ & Invisible & 2:1:1 \\
\hline
$\nu_3$ & $\nu_2$ & 1.4--1.6:1:1 \\
\hline
$\nu_3$ & $\nu_1$ & 2.4--2.8:1:1 \\
\hline
$\nu_3$ & 50\%\,$\nu_1$, 50\%\,$\nu_2$ & 2:1:1 \\
\hline
\end{tabular}
\caption{The neutrino flavor ratios predicted for a variety of
neutrino decay models with decay mode as indicated~\cite{bdecay}.}
\label{decaytable}
\end{table}

Of course this requires a substantial event rate. As shown in
Table~\ref{fluxestable} the standard cosmogenic neutrino flux is
expected to generate only about 0.7 quasi-horizontal shower events
over 10 years. This is certainly insufficient for making the precision
measurements needed to identify the effects of neutrino decay. For the
nominal Waxman-Bahcall flux, Auger is expected to detect about 2.2
quasi-horizontal events and about 48 Earth-skimming events in 10
yr. However, if the cosmic ray galactic--extragalactic transition
happens at around $10^9$~GeV~\cite{Berezinsky:2005cq}, then the
required proton luminosity in the extragalactic sources increases
significantly. Then Auger would detect as many as 21 quasi-horizontal
and 350 Earth-skimming events over 10 years. This corresponds to a
2$\sigma$ measurement of their ratio of $0.06 \pm 0.026$, which would
exclude anomalous flavor composition with a $\nu_e$ content greater
than $\phi_{\nu_e}:\phi_{\nu_{\mu}}:\phi_{\nu_{\tau}} \simeq 2.5:1:1$.
 
Other possibilities for altering neutrino flavor ratios have been
explored~\cite{bmeasure,lorentzcpt}. If Lorentz invariance is violated
through modification of the usual dispersion relation for neutrinos by
non-renormalizable operators induced by quantum gravity effects, then
the fraction of tau neutrinos may be
suppressed~\cite{lorentzcpt}. Auger would then observed the ratio of
Earth-skimmers to quasi-horizontal events to decrease from about 20 to
close to zero.

\section{Conclusions}
\label{conclusions}

Our knowledge of fundamental interactions has largely been limited to
the energies up to which collider experiments have been able to
probe. The Tevatron, currently operating at Fermilab, produces
collisions with a center-of-mass energy slightly below 2 TeV, while
the Large Hadron Collider, under construction at CERN, will reach 14
TeV. By contrast, a typical neutrino observed at the Pierre Auger
Observatory will have an energy of ${\cal O}(10^9)$ GeV, corresponding
to a neutrino-nucleon center-of-mass energy exceeding 40 TeV. Although
the number of collisions which will be observed (i.e. the beam
luminosity) is far below that of collider experiments, Auger and other
experiments sensitive to ultra-high energy cosmic neutrinos have in
principle the ability to provide unique information on new physics
beyond the reach of any planned accelerator.

In addition to this advantage, cosmic neutrinos have traveled over
very great distances before reaching Earth, thus their detection also
constitutes an extremely long-baseline oscillation experiment. Instead
of being limited to phenomena which occur over minuscule fractions of
a second, cosmic neutrinos provide an exceptional window into
phenomena only evident over cosmological scales of length or time.

The Pierre Auger Observatory is capable of detecting two primary
classes of neutrino induced events --- quasi-horizontal, deeply
penetrating showers and (slightly) upgoing showers induced by
Earth-skimming tau neutrinos. Used separately, the rates of such
events are of limited use in probing new physics; since the spectrum
of cosmic neutrinos is currently unknown, an event rate cannot by
itself be used to determine the neutrino-nucleon interaction
cross-section. However by combining these two classes of
neutrino-induced events, it becomes possible to make a crude
cross-section measurement. As this cross-section is increased
(decreased), the rate of quasi-horizontal showers increases
(decreases) accordingly, while by contrast, the rate of slightly
upgoing showers is reduced (enhanced) since Earth-skimming tau
neutrinos become absorbed more (less). Thus the ratio of
quasi-horizontal, deeply penetrating showers to slightly upgoing
showers provides a check of the behaviour of the neutrino-nucleon
cross-section at ultra-high energies. The details of such a
measurement, of course, depend on the energy dependence of such
interactions, as well as their inelasticity and other characteristics.

These two types of neutrino-induced events also provide the
opportunity to constrain the ratios of flavors present in the
ultra-high energy cosmic neutrino spectrum. If these neutrinos are
generated through the decay of charged pions (as they are in most
models), they will reach Earth in nearly equal quantities of each
flavor after oscillations are taken into account. A larger than
expected rate of quasi-horizontal, deeply penetrating showers, in
comparison to the slightly upgoing shower rate, would thus indicate a
suppression of the tau neutrino component in the ultra-high energy
cosmic neutrino spectrum, due, for example, to neutrino decay.

In this study, we have considered several specific models in which
either the neutrino-nucleon cross-section, or the ratio of cosmic
neutrino flavors, deviates substantially from the expectation of the
perturbative Standard Model. We have studied enhancements in the
neutrino-nucleon cross-section in models with low-scale gravity,
variously described as due to the exchange of Kaluza-Klein gravitons,
the production of microscopic black holes, and/or string resonances.
We have also considered increases in the cross-section due to
non-perturbative Standard Model electroweak instanton induced
processes, which in contrast do not lead to a decrease in the
inelasticity. Regarding flavor ratio measurements, we have discussed
several models of decaying neutrinos.

It is difficult to precisely delineate the reach of these techniques
as this depends on the unknown flux of cosmic neutrinos at the
energies to which Auger is sensitive. We have considered both the
``guaranteed'' cosmogenic flux which sets a lower bound and the
Waxman-Bahcall flux which sets an upper bound. Further observations of
ultra-high energy cosmic rays by Auger itself will help to pin down
the expected neutrino flux.

Over the next few years, the Pierre Auger Observatory may well
identify the world's first ultra-high energy neutrino event. We have
attempted to illustrate the exciting new possibilities for probing new
physics that will be opened up by such a detection.

\acknowledgements{We wish to thank Andreas Ringwald and Tom Weiler for
a critical reading of the manuscript and helpful comments. LAA is
partially supported by the US NSF grant PHY-0457004. TH is supported
by the US DoE grant DE-FG02-95ER40896 and by the Wisconsin Alumni
Research Foundation; he would also like to thank the Aspen Center for
Physics for hospitality. SS acknowledges a PPARC Senior Fellowship
(PP/C506205/1).}


\newpage
\begin{thebibliography}{99}

\bibitem{Abraham:2004dt}
J.~Abraham {\it et al.} [Pierre Auger Collaboration],
Nucl.\ Instrum.\ Meth.\ A {\bf 523}, 50 (2004).

\bibitem{auger}
P.~Sommers {\it et al.,} [Pierre Auger Collaboration],
29th International Cosmic Ray Conference, Pune, India (2005),
arXiv:astro-ph/0507150.

\bibitem{augersim}
K.~S.~Capelle, J.~W.~Cronin, G.~Parente and E.~Zas,
Astropart.\ Phys.\  {\bf 8}, 321 (1998).

\bibitem{amanda}
E.~Andres {\it et al.} [AMANDA Collaboration],
Astropart.\ Phys.\  {\bf 13}, 1 (2000).

\bibitem{Han:2004kq}
T.~Han and D.~Hooper,
New J.\ Phys.\  {\bf 6}, 150 (2004).

\bibitem{atten}
L.~A.~Anchordoqui, Z.~Fodor, S.~D.~Katz, A.~Ringwald and H.~Tu,
JCAP {\bf 0506} (2005) 013.

\bibitem{cross}
R.~Gandhi, C.~Quigg, M.~H.~Reno and I.~Sarcevic,
Phys.\ Rev.\ D {\bf 58}, 093009 (1998);
R.~Gandhi, C.~Quigg, M.~H.~Reno and I.~Sarcevic,
Astropart.\ Phys.\  {\bf 5}, 81 (1996).

\bibitem{Anchordoqui:2002vb}
L.~A.~Anchordoqui, J.~L.~Feng, H.~Goldberg and A.~D.~Shapere,
Phys.\ Rev.\ D {\bf 66}, 103002 (2002).

\bibitem{doublebang}
J.~G.~Learned and S.~Pakvasa,
Astropart.\ Phys.\  {\bf 3}, 267 (1995).

\bibitem{Glashow:W} S.~L.~Glashow,
Phys.\ Rev.\ {\bf 118}, 316 (1960).

\bibitem{tauevents}
J.~L.~Feng, P.~Fisher, F.~Wilczek and T.~M.~Yu,
Phys.\ Rev.\ Lett.\  {\bf 88}, 161102 (2002);
C.~Aramo, A.~Insolia, A.~Leonardi, G.~Miele, L.~Perrone, O.~Pisanti and D.~V.~Semikoz,
Astropart.\ Phys.\  {\bf 23}, 65 (2005);
S.~I.~Dutta, Y.~Huang and M.~H.~Reno,
Phys.\ Rev.\ D {\bf 72}, 013005 (2005).

\bibitem{regeneration}
F.~Halzen and D.~Saltzberg,
Phys.\ Rev.\ Lett.\  {\bf 81}, 4305 (1998).

\bibitem{zas}
E.~Zas,
New J.\ Phys.\  {\bf 7}, 130 (2005).

\bibitem{andes}
X.~Bertou, P.~Billoir, O.~Deligny, C.~Lachaud and A.~Letessier-Selvon,
Astropart.\ Phys.\  {\bf 17}, 183 (2002).

\bibitem{Berezinsky:1969}
V.~S.~Berezinsky and G.~T.~Zatsepin,
Phys.\ Lett.\ B {\bf 28} 423 (1969);
V.~S.~Berezinsky and A.~Y.~Smirnov,
Phys.\ Lett.\ B {\bf 48}, 269 (1974);
F.~W.~Stecker,
Astrophys.\ J.\  {\bf 228} (1979) 919.

\bibitem{gzk}
K.~Greisen,
Phys.\ Rev.\ Lett.\  {\bf 16}, 748 (1966);
G.~T.~Zatsepin and V.~A.~Kuzmin,
JETP Lett.\  {\bf 4}, 78 (1966)
[Pisma Zh.\ Eksp.\ Teor.\ Fiz.\ {\bf 4}, 114 (1966)].

\bibitem{Yoshida:pt}
S.~Yoshida and M.~Teshima,
Prog.\ Theor.\ Phys.\  {\bf 89}, 833 (1993);
R.~J.~Protheroe and P.~A.~Johnson,
Astropart.\ Phys.\  {\bf 4} (1996) 253.

\bibitem{engel}
R.~Engel, D.~Seckel and T.~Stanev,
Phys.\ Rev.\ D {\bf 64}, 093010 (2001).

\bibitem{Hill:1985mk}
C.~T.~Hill and D.~N.~Schramm,
Phys.\ Rev.\ D {\bf 31}, 564 (1985).

\bibitem{Bird:1993yi}
D.~J.~Bird {\it et al.}  [Fly's Eye Collaboration],
Phys.\ Rev.\ Lett.\  {\bf 71} (1993) 3401.

\bibitem{Bergman:2004bk}
D.~R.~Bergman  [HiRes Collaboration],
Nucl.\ Phys.\ Proc.\ Suppl.\  {\bf 136}, 40 (2004).

\bibitem{lowcrossover}
M.~Ahlers, L.~A.~Anchordoqui, H.~Goldberg, F.~Halzen, 
A.~Ringwald and T.~J.~Weiler,
Phys.\ Rev.\ D {\bf 72}, 023001 (2005).

\bibitem{Fodor:2003ph} 
Z.~Fodor, S.~D.~Katz, A.~Ringwald and H.~Tu,
JCAP {\bf 0311}, 015 (2003).

\bibitem{heavycosmogenic}
D.~Hooper, A.~Taylor and S.~Sarkar,
Astropart.\ Phys.\  {\bf 23}, 11 (2005);
M.~Ave, N.~Busca, A.~V.~Olinto, A.~A.~Watson and T.~Yamamoto,
Astropart.\ Phys.\  {\bf 23}, 19 (2005).

\bibitem{review}
F.~Halzen and D.~Hooper,
Rept.\ Prog.\ Phys.\  {\bf 65}, 1025 (2002).

\bibitem{wb}
E.~Waxman and J.~N.~Bahcall,
Phys.\ Rev.\ D {\bf 59}, 023002 (1999);
J.~N.~Bahcall and E.~Waxman,
Phys.\ Rev.\ D {\bf 64}, 023002 (2001).

\bibitem{equal}
L.~A.~Anchordoqui, H.~Goldberg, F.~Halzen and T.~J.~Weiler,
arXiv:hep-ph/0410003.

\bibitem{mpr}
J.~P.~Rachen, R.~J.~Protheroe and K.~Mannheim,
arXiv:astro-ph/9908031;
K.~Mannheim, R.~J.~Protheroe and J.~P.~Rachen,
Phys.\ Rev.\ D {\bf 63}, 023003 (2001).

\bibitem{hidden}
F.~W.~Stecker, C.~Done, M.~H.~Salamon and P.~Sommers,
Phys.\ Rev.\ Lett.\  {\bf 66}, 2697 (1991)
[Erratum-ibid.\  {\bf 69}, 2738 (1992)].

\bibitem{topdown}
V.~Berezinsky, M.~Kachelriess and A.~Vilenkin,
Phys.\ Rev.\ Lett.\  {\bf 79}, 4302 (1997);
M.~Birkel and S.~Sarkar,
Astropart.\ Phys.\  {\bf 9}, 297 (1998);
Z.~Fodor and S.~D.~Katz,
Phys.\ Rev.\ Lett.\  {\bf 86}, 3224 (2001);
S.~Sarkar and R.~Toldra,
Nucl.\ Phys.\ B {\bf 621}, 495 (2002);
C.~Barbot and M.~Drees,
Astropart.\ Phys.\  {\bf 20}, 5 (2003).

\bibitem{augerphoton}
M.~Risse  [Pierre Auger Collaboration],
arXiv:astro-ph/0507402.

\bibitem{dreesneutrinos}
C.~Barbot, M.~Drees, F.~Halzen and D.~Hooper,
Phys.\ Lett.\ B {\bf 555}, 22 (2003).

\bibitem{Gribov:1984tu}
L.~V.~Gribov, E.~M.~Levin and M.~G.~Ryskin,
Phys.\ Rept.\  {\bf 100}, 1 (1983);
A.~H.~Mueller and J.~W.~Qiu,
Nucl.\ Phys.\ B {\bf 268}, 427 (1986).

\bibitem{Kwiecinski:1990tb}
J.~Kwiecinski and A.~D.~Martin,
Phys.\ Rev.\ D {\bf 43}, 1560 (1991).

\bibitem{Arkani-Hamed:1998rs}
N.~Arkani-Hamed, S.~Dimopoulos and G.~R.~Dvali,
Phys.\ Lett.\ B {\bf 429}, 263 (1998).

\bibitem{Randall:1999ee}
L.~Randall and R.~Sundrum,
Phys.\ Rev.\ Lett.\  {\bf 83}, 3370 (1999).

\bibitem{Carena:1998gd}
M.~Carena, D.~Choudhury, S.~Lola and C.~Quigg,
Phys.\ Rev.\ D {\bf 58}, 095003 (1998).

\bibitem{Nussinov:1998jt}
S.~Nussinov and R.~Shrock,
Phys.\ Rev.\ D {\bf 59}, 105002 (1999).

\bibitem{Jain:2000pu}
P.~Jain, D.~W.~McKay, S.~Panda and J.~P.~Ralston,
Phys.\ Lett.\ B {\bf 484}, 267 (2000).

\bibitem{Feng:2001ib}
J.~L.~Feng and A.~D.~Shapere,
Phys.\ Rev.\ Lett.\  {\bf 88}, 021303 (2002);
L.~Anchordoqui and H.~Goldberg,
Phys.\ Rev.\ D {\bf 65}, 047502 (2002);
A.~Ringwald and H.~Tu,
Phys.\ Lett.\ B {\bf 525}, 135 (2002).

\bibitem{Domokos:1998ry}
G.~Domokos and S.~Kovesi-Domokos,
Phys.\ Rev.\ Lett.\  {\bf 82}, 1366 (1999).

\bibitem{measureair}
A.~Kusenko and T.~J.~Weiler,
Phys.\ Rev.\ Lett.\  {\bf 88}, 161101 (2002);
D.~Hooper,
Phys.\ Rev.\ D {\bf 65}, 097303 (2002).

\bibitem{Anchordoqui:1998nq}
L.~A.~Anchordoqui, M.~T.~Dova, L.~N.~Epele and S.~J.~Sciutto,
Phys.\ Rev.\ D {\bf 59}, 094003 (1999).

\bibitem{Anchordoqui:2005pn}
L.~A.~Anchordoqui, J.~L.~Feng and H.~Goldberg,
arXiv:hep-ph/0504228.

\bibitem{Eidelman:2004wy}
S.~Eidelman {\it et al.}  [Particle Data Group],
Phys.\ Lett.\ B {\bf 592}, 1 (2004).

\bibitem{Anchordoqui:2003vc}
L.~A.~Anchordoqui, H.~Goldberg, F.~Halzen and T.~J.~Weiler,
Phys.\ Lett.\ B {\bf 593}, 42 (2004).

\bibitem{Davoudiasl:1999jd} See e.g.,
H.~Davoudiasl, J.~L.~Hewett and T.~G.~Rizzo,
Phys.\ Rev.\ Lett.\  {\bf 84}, 2080 (2000);
S.~B.~Giddings and E.~Katz,
J.\ Math.\ Phys.\  {\bf 42}, 3082 (2001);
L.~A.~Anchordoqui, H.~Goldberg and A.~D.~Shapere,
Phys.\ Rev.\ D {\bf 66}, 024033 (2002);
A.~V.~Kisselev and V.~A.~Petrov,
Phys.\ Rev.\ D {\bf 71}, 124032 (2005);
A.~V.~Kisselev,
Eur.\ Phys.\ J.\ C {\bf 42}, 217 (2005);
D.~Stojkovic,
Phys.\ Rev.\ Lett.\  {\bf 94}, 011603 (2005).

\bibitem{Giudice:1998ck}
G.~F.~Giudice, R.~Rattazzi and J.~D.~Wells,
Nucl.\ Phys.\ B {\bf 544}, 3 (1999);
T.~Han, J.~D.~Lykken and R.~J.~Zhang,
Phys.\ Rev.\ D {\bf 59}, 105006 (1999).

\bibitem{Bando:1999di}
M.~Bando, T.~Kugo, T.~Noguchi and K.~Yoshioka,
Phys.\ Rev.\ Lett.\  {\bf 83}, 3601 (1999).

\bibitem{Anchordoqui:2000uh}
L.~Anchordoqui, H.~Goldberg, T.~McCauley, T.~Paul, S.~Reucroft 
and J.~Swain,
Phys.\ Rev.\ D {\bf 63}, 124009 (2001);
L.~Anchordoqui, T.~Paul, S.~Reucroft and J.~Swain,
Int.\ J.\ Mod.\ Phys.\ A {\bf 18}, 2229 (2003).

\bibitem{Kachelriess:2000cb}
M.~Kachelriess and M.~Plumacher,
Phys.\ Rev.\ D {\bf 62}, 103006 (2000),

\bibitem{Emparan:2001kf}
R.~Emparan, M.~Masip and R.~Rattazzi,
Phys.\ Rev.\ D {\bf 65}, 064023 (2002).

\bibitem{Illana:2005pu}
J.~I.~Illana, M.~Masip and D.~Meloni,
Phys.\ Rev.\ D {\bf 72}, 024003 (2005).

\bibitem{Anchordoqui:2004bd} 
Multiple KK interactions would produce a shower with characteristics
similar to those of a massive ($\sim 500$~GeV) hadron-induced shower:
L.~Anchordoqui, H.~Goldberg and C.~Nunez,
Phys.\ Rev.\ D {\bf 71}, 065014 (2005).

\bibitem{Banks:1999gd}
T.~Banks and W.~Fischler,
arXiv:hep-th/9906038;
S.~B.~Giddings and S.~Thomas,
Phys.\ Rev.\ D {\bf 65}, 056010 (2002);
S.~Dimopoulos and G.~Landsberg,
Phys.\ Rev.\ Lett.\  {\bf 87}, 161602 (2001).

\bibitem{Thorne:ji}
K.~S.~Thorne,
in {\it Magic Without Magic: John Archibald Wheeler}, edited by J. Klauder
(Freeman, San Francisco, 1972) p.231.

\bibitem{Myers:un}
R.~C.~Myers and M.~J.~Perry,
Annals Phys.\  {\bf 172}, 304 (1986);
P.~C.~Argyres, S.~Dimopoulos and J.~March-Russell,
Phys.\ Lett.\ B {\bf 441}, 96 (1998).

\bibitem{Eardley:2002re}
D.~M.~Eardley and S.~B.~Giddings,
Phys.\ Rev.\ D {\bf 66}, 044011 (2002).
See also
S.~B.~Giddings and V.~S.~Rychkov,
Phys.\ Rev.\ D {\bf 70}, 104026 (2004);
V.~S.~Rychkov,
Int.\ J.\ Mod.\ Phys.\ A {\bf 20}, 2398 (2005).

\bibitem{Yoshino:2002tx}
H.~Yoshino and Y.~Nambu,
Phys.\ Rev.\ D {\bf 67}, 024009 (2003).

\bibitem{D'Eath:hb} R.~Penrose, unpublished (1974);
P.~D.~D'Eath and P.~N.~Payne,
Phys.\ Rev.\ D {\bf 46}, 658 (1992);
Phys.\ Rev.\ D {\bf 46}, 675 (1992);
Phys.\ Rev.\ D {\bf 46}, 694 (1992).

\bibitem{Hawking:1975sw}
S.~W.~Hawking,
Commun.\ Math.\ Phys.\  {\bf 43}, 199 (1975).

\bibitem{Yoshino:2002br}
H.~Yoshino and Y.~Nambu,
Phys.\ Rev.\ D {\bf 66}, 065004 (2002).

\bibitem{Dimopoulos:2001qe}
S.~Dimopoulos and R.~Emparan,
Phys.\ Lett.\ B {\bf 526}, 393 (2002).

\bibitem{Antoniadis:1998ig}
I.~Antoniadis, N.~Arkani-Hamed, S.~Dimopoulos, G.~R.~Dvali,
Phys.\ Lett.\ B {\bf 436}, 257 (1998).
See also,  D.~Cremades, L.~E.~Ibanez and F.~Marchesano,
Nucl.\ Phys.\ B {\bf 643}, 93 (2002);
C.~Kokorelis,
Nucl.\ Phys.\ B {\bf 677}, 115 (2004).

\bibitem{Chamblin:2003wg} 
For the parameter space and energies of relevance to this paper,
accretion by the BH of surrounding particles is negligible:
A.~Chamblin, F.~Cooper and G.~C.~Nayak,
Phys.\ Rev.\ D {\bf 69}, 065010 (2004).

\bibitem{Frolov:2002xf}
V.~P.~Frolov and D.~Stojkovic,
Phys.\ Rev.\ D {\bf 67}, 084004 (2003);
Phys.\ Rev.\ D {\bf 68}, 064011 (2003);
V.~P.~Frolov, D.~V.~Fursaev and D.~Stojkovic,
JHEP {\bf 0406}, 057 (2004);
V.~P.~Frolov, D.~V.~Fursaev and D.~Stojkovic,
Class.\ Quant.\ Grav.\  {\bf 21}, 3483 (2004).

\bibitem{Han:2002yy}
T.~Han, G.~D.~Kribs and B.~McElrath,
Phys.\ Rev.\ Lett.\  {\bf 90}, 031601 (2003);
L.~Anchordoqui and H.~Goldberg,
Phys.\ Rev.\ D {\bf 67}, 064010 (2003).

\bibitem{Kanti:2002nr}
P.~Kanti and J.~March-Russell,
Phys.\ Rev.\ D {\bf 66}, 024023 (2002);
Phys.\ Rev.\ D {\bf 67}, 104019 (2003).

\bibitem{Emparan:2000rs}
R.~Emparan, G.~T.~Horowitz and R.~C.~Myers,
Phys.\ Rev.\ Lett.\  {\bf 85}, 499 (2000).

\bibitem{Harris:2003eg}
C.~M.~Harris and P.~Kanti,
JHEP {\bf 0310}, 014 (2003).

\bibitem{Cavaglia:2003hg}
M.~Cavaglia,
Phys.\ Lett.\ B {\bf 569}, 7 (2003).

\bibitem{Yoshino:2005hi}
H.~Yoshino and V.~S.~Rychkov,
Phys.\ Rev.\ D {\bf 71}, 104028 (2005).

\bibitem{greybody} These numbers were obtained by evaluating the
numerical results of Ref.~\cite{Harris:2003eg} at
$\langle Q \rangle,$ normalizing the cross-sections results to the
capture area $A_4$ defined in Eq.~(\ref{area}).

\bibitem{Preskill:1991tb}
J.~Preskill, P.~Schwarz, A.~D.~Shapere, S.~Trivedi and F.~Wilczek,
Mod.\ Phys.\ Lett.\ A {\bf 6}, 2353 (1991).

\bibitem{Hagedorn:st}
R.~Hagedorn,
Nuovo Cim.\ Suppl.\  {\bf 3}, 147 (1965).

\bibitem{Frautschi:1971ij}
S.~Frautschi,
Phys.\ Rev.\ D {\bf 3}, 2821 (1971);
R.~D.~Carlitz,
Phys.\ Rev.\ D {\bf 5}, 3221 (1972).

\bibitem{Hawking:de}
S.~W.~Hawking,
Phys.\ Rev.\ D {\bf 13}, 191 (1976).

\bibitem{Bowick:1985af}
M.~J.~Bowick, L.~Smolin and L.~C.~Wijewardhana,
Phys.\ Rev.\ Lett.\  {\bf 56}, 424 (1986).

\bibitem{Veneziano:1986zf}
G.~Veneziano,
Europhys.\ Lett.\  {\bf 2}, 199 (1986).

\bibitem{Susskind:ws}
L.~Susskind,
arXiv:hep-th/9309145.

\bibitem{Das:1996wn}
S.~R.~Das and S.~D.~Mathur,
Nucl.\ Phys.\ B {\bf 478}, 561 (1996);
Nucl.\ Phys.\ B {\bf 482}, 153 (1996);
J.~M.~Maldacena and A.~Strominger,
Phys.\ Rev.\ D {\bf 55}, 861 (1997).

\bibitem{Horowitz:1996nw}
G.~T.~Horowitz and J.~Polchinski,
Phys.\ Rev.\ D {\bf 55}, 6189 (1997);
T.~Damour and G.~Veneziano,
Nucl.\ Phys.\ B {\bf 568}, 93 (2000).

\bibitem{Amati:1999fv}
D.~Amati and J.~G.~Russo,
Phys.\ Lett.\ B {\bf 454}, 207 (1999).
Equivalence in the evaporation properties of highly excited strings and BHs has also been recently considered by
G.~Domokos and S.~Kovesi-Domokos,
arXiv:hep-ph/0307099.

\bibitem{Horowitz:1997jc}
G.~T.~Horowitz and J.~Polchinski,
Phys.\ Rev.\ D {\bf 57}, 2557 (1998).

\bibitem{Amati:1987wq}
D.~Amati, M.~Ciafaloni and G.~Veneziano,
Phys.\ Lett.\ B {\bf 197}, 81 (1987);
Int.\ J.\ Mod.\ Phys.\ A {\bf 3}, 1615 (1988).

\bibitem{Anchordoqui:2003jr}
L.~A.~Anchordoqui, J.~L.~Feng, H.~Goldberg and A.~D.~Shapere,
Phys.\ Rev.\ D {\bf 68}, 104025 (2003).

\bibitem{Pumplin:2002vw}
J.~Pumplin, D.~R.~Stump, J.~Huston, H.~L.~Lai, P.~Nadolsky and W.~K.~Tung,
JHEP {\bf 0207}, 012 (2002);
D.~Stump, J.~Huston, J.~Pumplin, W.~K.~Tung, H.~L.~Lai, S.~Kuhlmann 
and J.~F.~Owens,
hep-ph/0303013.

\bibitem{Frolov:2002as}
V.~P.~Frolov and D.~Stojkovic,
Phys.\ Rev.\ D {\bf 66}, 084002 (2002);
V.~P.~Frolov and D.~Stojkovic,
Phys.\ Rev.\ Lett.\  {\bf 89}, 151302 (2002).

\bibitem{Polchinski:1996na}
J.~Polchinski,
hep-th/9611050.
Here we consider the Type I relation $M_{10} = (8 \pi^5)^{1/8} \,
M_{\rm s}/g_{\rm s}^{1/4}$.

\bibitem{Anchordoqui:2003ug}
L.~A.~Anchordoqui, J.~L.~Feng, H.~Goldberg and A.~D.~Shapere,
Phys.\ Lett.\ B {\bf 594}, 363 (2004).

\bibitem{Anchordoqui:2004xb}
L.~Anchordoqui, M.~T.~Dova, A.~Mariazzi, T.~McCauley, T.~Paul, S.~Reucroft 
and J.~Swain,
Annals Phys.\  {\bf 314}, 145 (2004).

\bibitem{Cornet:2001gy}
F.~Cornet, J.~I.~Illana and M.~Masip,
Phys.\ Rev.\ Lett.\  {\bf 86}, 4235 (2001).

\bibitem{stringy2}
J.~Alvarez-Muniz, F.~Halzen, T.~Han and D.~Hooper,
Phys.\ Rev.\ Lett.\  {\bf 88}, 021301 (2002);
J.~J.~Friess, T.~Han and D.~Hooper,
Phys.\ Lett.\ B {\bf 547}, 31 (2002).

\bibitem{Kowalski:2002gb}
M.~Kowalski, A.~Ringwald and H.~Tu,
Phys.\ Lett.\ B {\bf 529}, 1 (2002)l;
J.~Alvarez-Muniz, J.~L.~Feng, F.~Halzen, T.~Han and D.~Hooper,
Phys.\ Rev.\ D {\bf 65}, 124015 (2002).

\bibitem{Dutta:2002ca}
S.~I.~Dutta, M.~H.~Reno and I.~Sarcevic,
Phys.\ Rev.\ D {\bf 66}, 033002 (2002);
E.~J.~Ahn, M.~Ave, M.~Cavaglia and A.~V.~Olinto,
Phys.\ Rev.\ D {\bf 68}, 043004 (2003);
V.~Cardoso, M.~C.~Espirito Santo, M.~Paulos, M.~Pimenta and B.~Tome,
arXiv:hep-ph/0405056;
A.~Cafarella, C.~Coriano and T.~N.~Tomaras,
JHEP {\bf 0506}, 065 (2005).

\bibitem{Anchordoqui:2001cg}
L.~A.~Anchordoqui, J.~L.~Feng, H.~Goldberg and A.~D.~Shapere,
Phys.\ Rev.\ D {\bf 65}, 124027 (2002).

\bibitem{Belavin:1975fg}
A.~A.~Belavin, A.~M.~Polyakov, A.~S.~Shvarts and Y.~S.~Tyupkin,
Phys.\ Lett.\ B {\bf 59}, 85 (1975).

\bibitem{'tHooft:1976fv}
G.~'t Hooft,
Phys.\ Rev.\ D {\bf 14}, 3432 (1976)
[Erratum-ibid.\ D {\bf 18} (1978) 2199];
G.~'t Hooft,
Phys.\ Rev.\ Lett.\  {\bf 37}, 8 (1976).

\bibitem{Klinkhamer:1984di} 
This energy sets the scale for non-perturbative baryon+lepton number 
$(B+L)$ violation in the SM:
F.~R.~Klinkhamer and N.~S.~Manton,
Phys.\ Rev.\ D {\bf 30}, 2212 (1984).

\bibitem{Aoyama:1986ej}
H.~Aoyama and H.~Goldberg,
Phys.\ Lett.\ B {\bf 188}, 506 (1987);
A.~Ringwald,
Nucl.\ Phys.\ B {\bf 330}, 1 (1990);
O.~Espinosa,
Nucl.\ Phys.\ B {\bf 343}, 310 (1990).

\bibitem{McLerran:1989ab}
L.~D.~McLerran, A.~I.~Vainshtein and M.~B.~Voloshin,
Phys.\ Rev.\ D {\bf 42}, 171 (1990);
S.~Y.~Khlebnikov, V.~A.~Rubakov and P.~G.~Tinyakov,
Nucl.\ Phys.\ B {\bf 350}, 441 (1991);
P.~B.~Arnold and M.~P.~Mattis,
Phys.\ Rev.\ D {\bf 42}, 1738 (1990).

\bibitem{Khoze:1990bm}
V.~V.~Khoze and A.~Ringwald,
Nucl.\ Phys.\ B {\bf 355}, 351 (1991).

\bibitem{Mattis:1991bj} For a review see {\it e.g},
M.~P.~Mattis,
Phys.\ Rept.\  {\bf 214}, 159 (1992).

\bibitem{Ringwald:2002sw}
A.~Ringwald,
Phys.\ Lett.\ B {\bf 555}, 227 (2003);
F.~Bezrukov, D.~Levkov, C.~Rebbi, V.~A.~Rubakov and P.~Tinyakov,
Phys.\ Rev.\ D {\bf 68}, 036005 (2003);
F.~Bezrukov, D.~Levkov, C.~Rebbi, V.~A.~Rubakov and P.~Tinyakov,
Phys.\ Lett.\ B {\bf 574}, 75 (2003);
A.~Ringwald,
JHEP {\bf 0310}, 008 (2003).

\bibitem{Fodor:2004tr}
Z.~Fodor, S.~D.~Katz, A.~Ringwald and H.~Tu,
Phys.\ Lett.\ B {\bf 561}, 191 (2003).

\bibitem{Morris:1993wg}
D.~A.~Morris and A.~Ringwald,
Astropart.\ Phys.\  {\bf 2}, 43 (1994).

\bibitem{Ahlers:2005zy}
M.~Ahlers, A.~Ringwald and H.~Tu,
arXiv:astro-ph/0506698.

\bibitem{Han:2003ru}
T.~Han and D.~Hooper,
Phys.\ Lett.\ B {\bf 582}, 21 (2004).

\bibitem{Schechter:1981cv}
J.~Schechter and J.~W.~F.~Valle,
Phys.\ Rev.\ D {\bf 25}, 774 (1982).

\bibitem{Bahcall:1986gq}
J.~N.~Bahcall, S.~T.~Petcov, S.~Toshev and J.~W.~F.~Valle,
Phys.\ Lett.\ B {\bf 181}, 369 (1986);
J.~F.~Beacom and N.~F.~Bell,
Phys.\ Rev.\ D {\bf 65}, 113009 (2002).

\bibitem{bdecay}
J.~F.~Beacom, N.~F.~Bell, D.~Hooper, S.~Pakvasa and T.~J.~Weiler,
Phys.\ Rev.\ Lett.\  {\bf 90}, 181301 (2003);
J.~F.~Beacom, N.~F.~Bell, D.~Hooper, S.~Pakvasa and T.~J.~Weiler,
Phys.\ Rev.\ D {\bf 69}, 017303 (2004).

\bibitem{Anchordoqui:2005gj}
L.~A.~Anchordoqui, H.~Goldberg, M.~C.~Gonzalez-Garcia, F.~Halzen, 
D.~Hooper, S.~Sarkar and T.~J.~Weiler,
Phys. Rev. D {\bf 72}, 065019 (2005).

\bibitem{Berezinsky:2005cq}
V.~Berezinsky, A.~Z.~Gazizov and S.~I.~Grigorieva,
Phys.\ Lett.\ B {\bf 612}, 147 (2005).

\bibitem{bmeasure}
J.~F.~Beacom, N.~F.~Bell, D.~Hooper, S.~Pakvasa and T.~J.~Weiler,
Phys.\ Rev.\ D {\bf 68}, 093005 (2003);
J.~F.~Beacom, N.~F.~Bell, D.~Hooper, J.~G.~Learned, 
S.~Pakvasa and T.~J.~Weiler,
Phys.\ Rev.\ Lett.\  {\bf 92}, 011101 (2004);
D.~Hooper, D.~Morgan and E.~Winstanley,
Phys.\ Lett.\ B {\bf 609}, 206 (2005).

\bibitem{lorentzcpt}
D.~Hooper, D.~Morgan and E.~Winstanley,
Phys.\ Rev.\ D {\bf 72}, 065009 (2005).

\end{thebibliography}
\end{document}